\documentclass[preprint,12pt]{elsarticle}
\usepackage{float}
\usepackage{psfrag}
\usepackage{amsmath}
\usepackage{amsfonts}
\usepackage{graphicx}
\usepackage{color}
\usepackage[utf8]{inputenc}
\usepackage{lscape}
\usepackage{multicol}
\usepackage{srcltx}
\usepackage{overpic}
\usepackage{titletoc}
\usepackage{titlesec}
\usepackage{multirow}
\usepackage{tikz}
\usepackage{enumitem}
\usepackage[papersize={216mm,279mm},tmargin=20mm,bmargin=30mm,lmargin=30mm,rmargin=30mm]{geometry}

\usepackage{hyperref}
\usepackage{xcolor}
\usepackage{booktabs}
\usepackage{fancyhdr}
\usepackage{titlesec}
\usepackage{tcolorbox}
\usepackage{doi}
\usepackage{float}
\usepackage{caption}

\newcommand{\dif}{\mathrm d}

\journal{J. Phys.: Cond. Matter}

\begin{document}
\markboth{E. Mamani and D. Sanjin\'es}{effective interactions in a rectangular lattice}

\title{Long-range effective interactions in a rectangular tight-binding lattice within the semiclassical approximation}
\author{E. Mamani, D. Sanjin\'es}
\address{Carrera de F\'isica, Universidad Mayor de San Andr\'es, c. 27 Cota - Cota, casilla de correos 8635, La Paz, Bolivia.\\email: diegosanjinescastedo@gmail.com}

\begin{abstract}
We extend the semiclassical formalism of Dunlap--Kenkre~\cite{dunlap1986} and of Mamani \emph{et al.}~\cite{mamani2017} to obtain the effective Hamiltonian of a rectangular tight-binding lattice with interactions to all neighbors, in the presence of a rapidly oscillating field $\mathbf f(\omega t)$ and a static potential $V(x,y)$, both arbitrary, by means of: a canonical transformation, the semiclassical approximation and the Kapitza-type time-averaging techniques \cite{mamani2017}. At zeroth order in $1/\omega$, a field with even symmetry reparameterizes the hopping elements through factors in terms of Bessel functions. At order $1/\omega^2$, the induced hopping elements arise from the coupling between the oscillating field and the curvature of the static potential (its 2nd derivative). For fields $\mathbf f(\omega t)$ with odd symmetry, in addition to renormalizing the interactions, an effective band shift appears. For the two-dimensional lattice, we obtain the generalization of the exact dynamic localization condition of Dignam \emph{et al.}~\cite{dignam2002} and Domachuk \emph{et al.}~\cite{domachuk2002}, and we show that the resulting band-collapse is in general anisotropic --axial interactions are suppressed but some diagonal interactions persist--, so that exact dynamic localization in the stroboscopic sense is guaranteed only when the band-collapse is complete for all neighbors. Finally, we numerically identify an effective-duplication scenario for the rectangular lattice, associated with approximate solutions of a Pythagorean-type relation among zeros of the zeroth-order Bessel function.
\end{abstract}

\maketitle

\section{Introduction}
The study of electronic dynamics in solid-state lattices has attracted renewed interest, driven both by advances in experimental techniques for controlling electronic evolution and by the development of photonic platforms capable of emulating phenomena characteristic of condensed-matter systems~\cite{ablowitz2022,borsch2023,zhan2024}. Tight-binding Hamiltonians provide a standard framework for describing transport and localization in one- and two-dimensional lattices~\cite{karamlou2022}. Among the studied effects, those arising from the application of an external, homogeneous alternating-current (AC) electric field are relevant to quantum transport phenomena, such as dynamic localization and coherent control of tunneling, as well as to the manipulation of atomic motion in optical lattices and the control of the effective properties of periodically driven systems~\cite{weitenberg2021,shevchenko2023,eckardt2017}. Dynamic localization of a charged particle under an AC field was established by Dunlap and Kenkre~\cite{dunlap1986} and has been extended to other driving configurations, such as fields with a definite relative phase~\cite{singh2012}, simultaneous electric and magnetic fields~\cite{papp2008}, and synthetic lattices~\cite{wang2022}.

If, in addition, the AC fields are of high frequency, time-averaging methods can be applied to describe the effective dynamics of the system, following the approach introduced by Kapitza in the study of the inverted pendulum. This scheme has been extended to the analysis of periodically driven systems, making it possible to obtain time-independent effective Hamiltonians for a tight-binding lattice and to study phenomena of dynamical stabilization and modification of the system's properties \cite{bukov2015,eckardt2015}. This approach is related to the description of rapidly oscillating potentials~\cite{rahav2003a} and to Floquet effective Hamiltonians and high-frequency expansions in periodically driven systems~\cite{zhou2023}; in this context, conditions have been identified under which couplings between distant neighbors appear~\cite{itin2014}. The application of high-frequency AC fields makes it possible to obtain an effective Hamiltonian whose hopping terms can be controlled through the parameters of the external field. In particular, it is possible to suppress the dominant interactions between nearest neighbors and make the contributions of more distant neighbors relevant, which allows the dynamical properties of the lattice to be modified~\cite{mamani2017}. On the other hand, the semiclassical approximation makes it possible to describe the average dynamics of a particle in a tight-binding lattice by means of a time-independent effective potential, even in the presence of a spatially arbitrary static potential~\cite{martinez2014}, and constitutes the basis of Floquet engineering of bands and topological phases, both in effective models and in realistic materials~\cite{zhan2024}, in two-dimensional lattices driven by two frequencies~\cite{wang2023,ikeda2022}, in graphene-type lattices~\cite{zhang2023}, and in ultracold-atom quantum simulators~\cite{wangreview2026}.

In the presence of an electromagnetic field, the interaction of a charged particle with the field modifies the phase of the hopping amplitudes between lattice sites. In the tight-binding model, this coupling is incorporated by means of the Peierls substitution, which introduces a phase factor dependent on the electromagnetic vector potential into the hopping elements~\cite{zhan2024}. If the field is sinusoidal, the Peierls phase is also periodic, and upon averaging over a cycle, the effective hopping depends on the amplitude and frequency of the field. In the simplest case, that of nearest neighbors, the effective hopping element is proportional to ${J}_{0}(K)$, where ${J}_{0}$ is the zeroth-order Bessel function and $K$ is a dimensionless amplitude proportional to the amplitude of the electric field and inversely proportional to its frequency (see also the renormalization of hoppings in the high-frequency regime in Ref.~\cite{qcrossover2020}). The effective hopping can be reduced until it vanishes, giving rise to dynamic localization; Dunlap and Kenkre showed that this occurs for the values of $K$ that are zeros of ${J}_{0}$~\cite{dunlap1986}. When ${J}_{0}(K)$ changes sign, the effective hopping acquires a relative phase of $\pi$, which does not imply the appearance of an arbitrary complex phase. On the other hand, an extended tight-binding Hamiltonian, with kinetic energy of the form $2\sum\limits_{n}^{}{A}_{n}\cos (nap)$, can be traced back to the pioneering work of Dunlap and Kenkre~\cite{dunlap1986}, whose results on dynamic localization extend to long-range interactions.

In a rectangular lattice, the field may have components ${E}_{x}(t)$ and ${E}_{y}(t)$. If the hopping amplitudes or the lattice constants differ in the two directions, the field can renormalize the couplings along $x$ and $y$ differently, generating anisotropy in the dispersion and in transport. A field oriented solely along $x$ mainly modifies the hopping elements along $x$, while purely transverse hoppings along $y$ do not acquire a Peierls phase from that field. For a field with elliptical or circular polarization, of the form ${E}_{x}(t)={E}_{0x}\cos (\omega t)$, ${E}_{y}(t)={E}_{0y}\cos (\omega t+\delta )$, where the parameter $\delta$ is the relative phase between the two components. When $\delta \ne 0,\pi$, the field traces closed trajectories in the plane and breaks certain time symmetries, so that complex effective hopping amplitudes and direction-dependent phases can appear. Control of the relative phase and of the frequency mixing also modifies the conditions for dynamic localization~\cite{singh2012,papp2008,wang2023}.

Within this framework, Mamani \emph{et al.}~\cite{mamani2017}, using the semiclassical model, obtained the effective Hamiltonian comprising interactions up to first and second neighbors in the presence of an arbitrary static field $V(x)$ and a uniform rapidly oscillating field. The appearance of an interaction between first and second neighbors as a consequence of combining a rapidly oscillating field with a quadratic static potential was derived by applying the Kapitza averaging technique to a tight-binding Hamiltonian~\cite{mamani2017,martinez2014}; the semiclassical model leads more simply and expeditiously to the same conclusions as the strict quantum treatment, specifically in the study of the combined effects on electronic motion in the presence of fields that are arbitrarily dependent on space and time. In two dimensions, Mamani, Calcina-Nogales, and Sanjinés Castedo~\cite{mamani2021} considered a square lattice with interactions up to first and second neighbors, an arbitrary static potential, and a homogeneous high-frequency electric field, and constructed, by means of the semiclassical approximation and Kapitza averaging, an effective Hamiltonian up to order ${\omega }^{-2}$ (their work is also related to an LC circuit of two meshes with discrete charge). Schematically, the effective Hamiltonian has the form ${H}_{eff}={H}_{eff}^{(0)}+1/{\omega }^{2}{H}_{eff}^{(2)}+O({\omega }^{-3})$, where the orders effectively present depend on the rapidly oscillating field and on the static potential. The authors show that the field parameters allow the first-neighbor interactions to be suppressed while retaining the second-neighbor ones, and that at order ${\omega }^{-2}$ third-neighbor contributions appear~\cite{mamani2021}.

Exact dynamic localization (EDL) in the presence of long-range interactions requires more than the nearest-neighbor criterion. Dignam and de Sterke~\cite{dignam2002} established conditions for EDL in a general single-band model and showed that, with long-range interactions, it is not enough to apply that criterion directly: conditions on the field are required that guarantee the cancellation of the entire effective dynamics, not merely the renormalization of a particular hopping. Domachuk \emph{et al.}~\cite{domachuk2002} analyzed the waveforms capable of inducing dynamic localization: in the nearest-neighbor limit, symmetric AC fields can satisfy the conditions for certain amplitudes, but in the general model EDL requires more restrictive conditions. García Foronda, Sanjinés Castedo, and Mamani Carlo~\cite{garcia2021} studied a one-dimensional lattice with interactions to all neighbors under a homogeneous rapidly oscillating field, and identified EDL with the vanishing of the effective terms ${G}_{0}(n)=0$ for all $n\ne 0$, further comparing the semiclassical formalism with the quantum results of Dignam and de Sterke; they found that symmetric triangular and sawtooth waves do not simultaneously satisfy these conditions, whereas a rectangular wave does admit solutions. Martínez-Quintana and González-Díaz~\cite{mqg2017} examined the conditions for dynamic localization in an effective tight-binding Hamiltonian under a homogeneous high-frequency electric field using the Magnus--Floquet method. In two- and three-dimensional lattices, Nazareno and Lépine~\cite{nazareno1997} point out that the orientation of the electric field is crucial for wavefunction confinement, although their analysis does not constitute a general proof of band collapse for an arbitrary two-dimensional lattice with hoppings to all neighbors.

From the cited works it follows that the two-dimensional results with a rapidly oscillating field are limited to interactions up to first and second neighbors~\cite{mamani2021}, whereas the treatment with interactions to all neighbors and EDL has been formulated in one dimension~\cite{garcia2021}. In two dimensions it is additionally necessary to show that the chosen field allows the cancellation conditions to be satisfied for all hopping elements of the lattice: the one-dimensional theorem of Dignam and de Sterke does not by itself constitute a proof of EDL in a two-dimensional lattice with long-range interactions. It is also worth distinguishing three situations: suppressing some hopping elements, collapsing the band, and achieving EDL. A band may change its dispersion or narrow without a complex phase appearing in the Hamiltonian.

The relevance of the problem lies in the possibility of redesigning the connectivity of the effective lattice through interaction engineering. As an illustration, consider a square lattice of lattice constant $a$ with nearest-neighbor interactions $(\pm a,0)$ and $(0,\pm a)$, axial second-neighbor interactions $(\pm 2a,0)$ and $(0,\pm 2a)$, and diagonal nearest-neighbor interactions $(\pm a,\pm a)$. For real, symmetric amplitudes, the initial dispersion has the form ${E}_{0}(k)=-2{A}_{1}[\cos (a{k}_{x})+\cos (a{k}_{y})]-2{A}_{2}[\cos (2a{k}_{x})+\cos (2a{k}_{y})]-4{A}_{d}\cos (a{k}_{x})\cos (a{k}_{y})$, where ${A}_{1}$ is the nearest-neighbor hopping element, ${A}_{2}$ the second-neighbor hopping element, and ${A}_{d}$ the diagonal hopping. If interaction engineering could achieve ${A}_{1}^{eff}=0$, ${A}_{d}^{eff}=0$, and ${A}_{2}^{eff}\ne 0$, the dominant dispersion would take the form ${E}_{eff}(k)=-2{A}_{2}^{eff}[\cos (2a{k}_{x})+\cos (2a{k}_{y})]$, a band with effective spatial periodicity $2a$ in both directions. Strictly speaking, if only the axial hoppings of length $2a$ survive, the lattice splits into four decoupled sublattices, labeled by the parity of the integer coordinates, each with lattice constant $2a$; if the effective hopping lengths along $x$ and $y$ differ, each sublattice is rectangular, with primitive vectors $(2a,0)$ and $(0,2b)$. This description corresponds to the dominant Hamiltonian ${H}_{eff}^{(0)}$: a residual third-neighbor axial hopping, of length $3a$, connects sites of different parity and, if its amplitude is nonzero, recouples the sublattices. For this reason, the connectivity defined by ${H}_{eff}^{(0)}$, the residual terms of ${H}_{eff}^{(2)}$, and the time scale of transitions between sublattices must be analyzed separately.

This consideration is tied to the role of the different orders of the expansion in the effective Hamiltonian. In the generic regime, the zeroth-order hopping elements do not vanish and constitute the dominant term, so that the order-${\omega }^{-2}$ corrections are small; conversely, if the zeroth-order terms vanish under some hypothesis, the order-${\omega }^{-2}$ terms are the induced terms and determine the residual dynamics, and so must not be automatically discarded. In both cases, the zeroth-order terms are affected by the hopping elements renormalized due to the presence of the rapidly oscillating field. When such a field is applied to a two-dimensional lattice with interactions to all neighbors, there exists the possibility that, for certain parameter values, the renormalized hopping elements vanish completely, which would amount to a freezing of the band and, eventually, to EDL in the sense of Dignam and de Sterke and of Domachuk \emph{et al.}~\cite{dignam2002,domachuk2002}. Likewise, when only a high-frequency electric field is applied, a band shift is not ruled out, depending on the even or odd symmetry of the waveform: the usual tight-binding band structure, $-A\cos (a{k}_{x})$, could take the form $-A'\cos (a{k}_{x}+\phi )$, where $A'$ is the renormalized hopping element and $\phi$ the shift. The available results on Floquet engineering in two-dimensional lattices~\cite{wang2023,zhang2023} do not automatically show that a single-frequency field produces such a shift, so this point requires specific analysis.

The purpose of this work is to extend the calculations reported in Ref.~\cite{mamani2017} to an extended tight-binding Hamiltonian for a rectangular lattice that includes interaction with \emph{all} neighbors, in the presence of an arbitrary static field $V(x,y)$ and a uniform rapidly oscillating field with arbitrary time dependence. The time-independent effective Hamiltonian will be derived from the time-dependent Hamiltonian, applying the semiclassical approximation and the Kapitza time-averaging method (separation of motions) up to terms of order ${\omega }^{-2}$.

The methodology therefore consists in applying the Kapitza time-averaging method to a tight-binding model extended to \emph{all} neighbors within the semiclassical approximation. The article is organized as follows. In Sec.~2 the theoretical model is set out and the tight-binding Hamiltonian is formulated, incorporating the interaction with \textit{all} neighbors for a rectangular lattice in the presence of a uniform rapidly oscillating field and an arbitrary static field; in Sec.~3 we start from Hamilton's equations and apply a type of canonical transformation; in Sec.~4 the Kapitza separation-of-motions method is applied; in Sec.~5 a Fourier series expansion of the rapidly oscillating electric field is carried out; in Sec.~6 the equations of motion for the slow position and momentum coordinates are obtained; in Sec.~7 the effective Hamiltonian $H(R,P)$ is obtained; in Sec.~8 the results and discussion of the cases of interest are presented, applying the general effective Hamiltonian up to order ${\omega }^{-2}$, and the main conclusions drawn from it are developed; and in Sec.~9 the most important final conclusions and some perspectives of interest are presented.
\section{Theoretical model and formulation of the Hamiltonian}

Consider the extended tight-binding Hamiltonian for a rectangular lattice, with lattice constants $a$ and $b$ along the $x$ and $y$ directions respectively, electric charge $e=1$ and $\hbar=1$, in the presence of an arbitrary static potential $V(x,y)$ and a uniform rapidly oscillating field of the form $xf_x(\omega t)+yf_y(\omega t)$ with a high frequency $\omega$:
\begin{equation}\label{eq:H}
H\left(x,y,k_x,k_y;t\right)=-\sum_{n,m}A_{n,m}\cos{\left(nak_x\right)}\cos{\left(mbk_y\right)}+xf_x\left(\omega t\right)+yf_y\left(\omega t\right)+V(x,y),
\end{equation}
where the sum over $n$ and $m$ runs over all integers except for the case $n=m=0$, and $A_{n,m}$ is the hopping element corresponding to the lattice vector $\mathbf{G}_{n,m}=na\;\mathbf{i}+mb\;\mathbf{j}$. We further assume that the hopping elements satisfy the reflection symmetry: $A_{\pm n,0}=A_{n}$ is the hopping element along the $x$ direction, $A_{0,\pm m}=B_{m}$ is the hopping element along the $y$ direction, and $A_{\pm n,\pm m}=C_{n,m}$ is the diagonal hopping element in the $xy$ plane.

\begin{figure}[h!]
\centering
\includegraphics[scale=0.4]{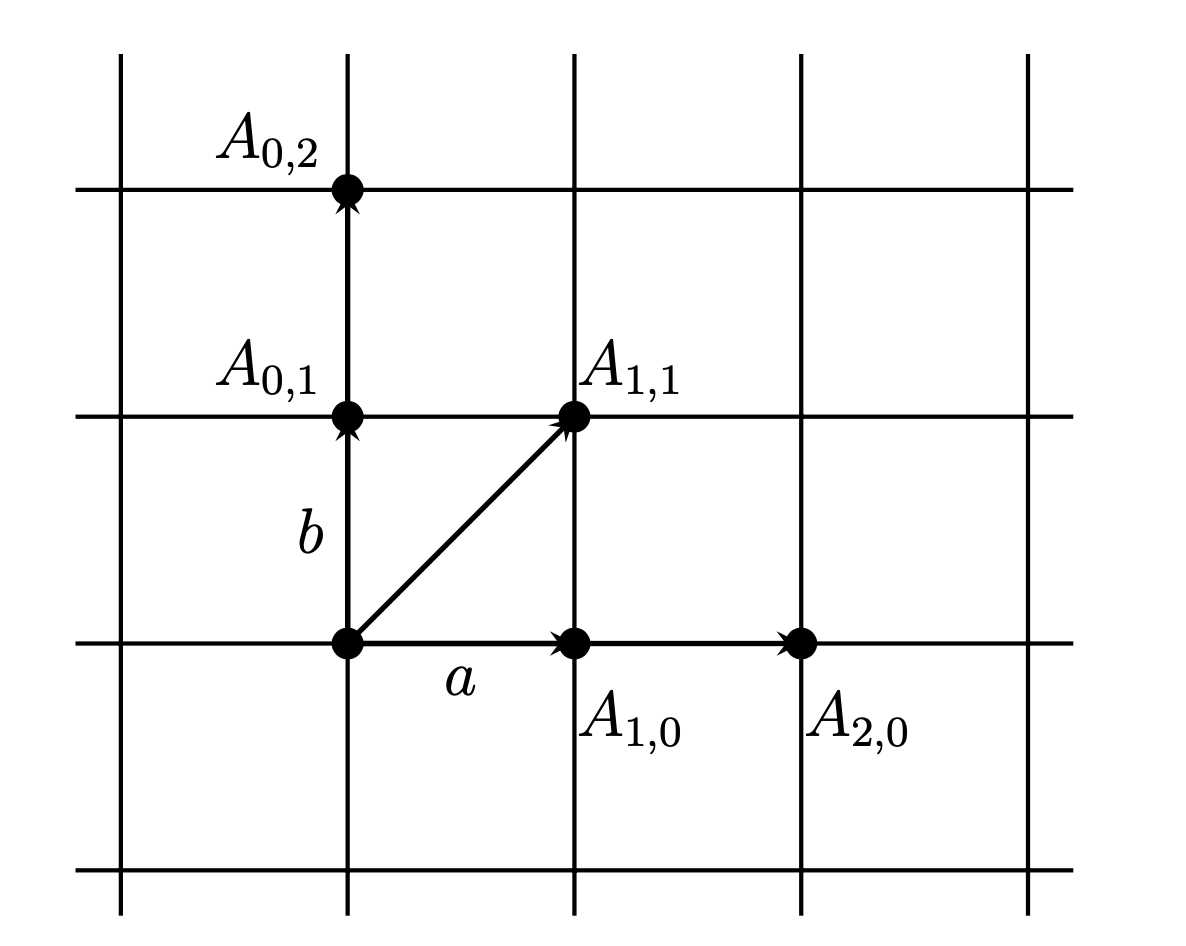}
\caption{Rectangular lattice with first- and second-neighbor interactions in the horizontal and vertical directions, and with a diagonal interaction.}
\label{fig:red-rectangular}
\end{figure}

Under the symmetry of the hopping elements described above, Eq.~\eqref{eq:H} can also be written in the form:
\begin{equation}\label{eq:H1}
H\left(x,y,k_x,k_y;t\right)=-\Re\left(\sum_{n,m}A_{n,m}e^{i(nak_x+mbk_y)}\right)+xf_x\left(\omega t\right)+yf_y\left(\omega t\right)+V(x,y),
\end{equation}

\section{Hamilton's equations and canonical transformation}

From Hamilton's equations:
\begin{align}
  &\dot{x}= \frac{\partial H}{\partial k_x},
  \qquad
  &\dot{k}_x= -\frac{\partial H}{\partial x}, \label{eq:Hamilton1}\\
  &\dot{y}= \frac{\partial H}{\partial k_y},
  \qquad
  &\dot{k}_y= -\frac{\partial H}{\partial y},\label{eq:Hamilton2}
\end{align}
the equations of motion for the canonical variables $x$, $k_x$, $y$, and $k_y$, according to \eqref{eq:Hamilton1} and \eqref{eq:Hamilton2} for the Hamiltonian \eqref{eq:H1}, are:
\begin{align}
\dot{x}=&-\Re\left(\sum_{n,m}inaA_{n,m}e^{i(nak_x+mbk_y)}\right),\label{eq:x1}\\
\dot{k}_x=&-f_x\left(\omega t\right)-\partial_{x}V,\label{eq:k1}\\
\dot{y}=&-\Re\left(\sum_{n,m}imbA_{n,m}e^{i(nak_x+mbk_y)}\right),\label{eq:y1}\\
\dot{k}_y=&-f_y\left(\omega t\right)-\partial_{y}V.\label{eq:k2},
\end{align}
where $\partial_{x}V\equiv\partial V/\partial x$, $\partial_{y}V\equiv\partial V/\partial y$.

We define $\mathbf{g}(t)\equiv\int^{t}_{0}\mathbf{f}(\omega t')dt'$, with $\mathbf{f}(\omega t)=f_x(\omega t)\mathbf{i}+f_y(\omega t)\mathbf{j}$, and the canonical transformation:
\begin{align}
\mathbf{k}(t)=\mathbf{K}(t)-\mathbf{g}(t)\;\Rightarrow\;\dot{\mathbf{k}}(t)=\dot{\mathbf{K}}(t)-\mathbf{f}(\omega t),\label{eq:tc}
\end{align}
which is indeed canonical because it generates a translation in momentum space that does not depend on the coordinates $(x,y)$ or on $\mathbf{K}$. The old and new variables are $x=x'$, $y=y'$, $k_x=K_x-g_x(t)$, $k_y=K_y-g_y(t)$, and the Jacobian matrix is $J=\frac{\partial(x,y,k_x,k_y)}{\partial(x',y',K_x,K_y)}$; since $g_x(t)$, $g_y(t)$ do not depend on $K_x$, $K_y$, $x'$, or $y'$, the Jacobian matrix is the identity. This is equivalent to shifting the origin from which momentum is measured: previously it was measured relative to the origin, and now it is measured relative to the origin $\mathbf{g}(t)$; since only the origin is shifted and phase space is not deformed, areas remain unchanged, which is precisely the defining property of a canonical transformation.

Substituting \eqref{eq:tc} into \eqref{eq:x1}, \eqref{eq:k1}, \eqref{eq:y1}, and \eqref{eq:k2}, and using the vector notation $\mathbf{r}=x\mathbf{i}+y\mathbf{j}$, $\mathbf{G}=na\mathbf{i}+mb\mathbf{j}$, $\mathbf{k}=k_x\mathbf{i}+k_y\mathbf{j}$, and the identity $-i=e^{-i\pi/2}$, we obtain:
\begin{align}
\dot{\mathbf{r}}=&\Re\left(\sum_{n,m}\mathbf{G}A_{n,m}e^{i\left[\mathbf{G}\cdot\left(\mathbf{K}-\mathbf{g}\right)-\pi/2\right]}\right),\label{eq:rn}\\
\dot{\mathbf{K}}=&-\mathbf{\nabla}V,\label{eq:kn}
\end{align}

\section{Separation of motions: slow and fast coordinates}

We introduce the fast time $\tau=\omega t$ and propose, as a solution to the system of equations \eqref{eq:rn}, \eqref{eq:kn}, the decompositions:
\begin{align}
\mathbf{r}(t)&=\mathbf{R}(t)+\boldsymbol{\xi}(\tau),\label{eq:rs}\\
\mathbf{K}(t)&=\mathbf{P}(t)+\boldsymbol{\eta}(\tau),\label{eq:ks}
\end{align}
where $\mathbf{R}(t)$, $\mathbf{P}(t)$ are the slow coordinates and $\boldsymbol{\xi}(\tau)$, $\boldsymbol{\eta}(\tau)$ the fast coordinates. Differentiating \eqref{eq:rs}, \eqref{eq:ks} with respect to $t$, with d$\tau/$d$t=\omega$,
\begin{align}
\dot{\mathbf{r}}&=\dot{\mathbf{R}}+\omega\frac{\text{d}\boldsymbol{\xi}}{\text{d}\tau},\label{eq:rsd}\\
\dot{\mathbf{K}}&=\dot{\mathbf{P}}+\omega\frac{\text{d}\boldsymbol{\eta}}{\text{d}\tau},\label{eq:ksd}
\end{align}
substituting \eqref{eq:rsd}, \eqref{eq:ksd} into \eqref{eq:rn}, \eqref{eq:kn}:
\begin{align}
\dot{\mathbf{R}}+\omega\frac{d\boldsymbol{\xi}}{d\tau}=&\Re\left(\sum_{n,m}\mathbf{G}A_{n,m}e^{i\left[\mathbf{G}\cdot\left(\mathbf{P}+\boldsymbol{\eta}-\mathbf{g}\right)-\pi/2\right]}\right),\label{eq:rnn}\\
\dot{\mathbf{P}}+\omega\frac{d\boldsymbol{\eta}}{d\tau}=&-\mathbf{\nabla}V(\mathbf{R}+\boldsymbol{\xi}),\label{eq:knn}
\end{align}
we further use the usual definition for the time average $\langle\cdot\rangle=(1/T)\int^{T}_{0}(\cdot)\text{d}t$, where $T=2\pi/\omega$ is the period and $(\cdot)$ is the function being averaged. We assume that the fast coordinates have zero average over a period $T$ (which corresponds to the period of the oscillating field $\mathbf{f}(\omega t)$), while the slow coordinates vary very little, that is:
\begin{align}
\langle\boldsymbol{\xi}(\tau)\rangle&=0,\;\;\;&\langle\boldsymbol{\eta}(\tau)\rangle&=0,,\label{eq:corrap}\\
\langle\mathbf{R}(t)\rangle&=\mathbf{R}(t),\;\;\;&\langle\mathbf{P}(t)\rangle&=\mathbf{P}(t),,\label{eq:corlen}
\end{align}
taking the time average of the system of equations \eqref{eq:rnn}, \eqref{eq:knn} and using $\left\langle\text{d}\boldsymbol{\xi}/\text{d}\tau\right\rangle=\left\langle\text{d}\boldsymbol{\eta}/\text{d}\tau\right\rangle=\mathbf{0}$, since $\boldsymbol{\xi}$, $\boldsymbol{\eta}$ are periodic functions with zero average, so that the derivative of a periodic function also averages to zero over a full period, we obtain:
\begin{align}
\dot{\mathbf{R}}=&\Re\left(\sum_{n,m}\mathbf{G}A_{n,m}\left\langle e^{i\left[\mathbf{G}\cdot\left(\mathbf{P}+\boldsymbol{\eta}-\mathbf{g}\right)-\pi/2\right]}\right\rangle\right),\label{eq:rnn1}\\
\dot{\mathbf{P}}=&-\langle\mathbf{\nabla}V(\mathbf{R}+\boldsymbol{\xi})\rangle,\label{eq:knn1}
\end{align}
this system of equations constitutes the \textit{``effective equations of motion for the slow coordinates''}. On the other hand, subtracting \eqref{eq:rnn1} from \eqref{eq:rnn} and \eqref{eq:knn1} from \eqref{eq:knn} we obtain:
\begin{align}
\omega\frac{d\boldsymbol{\xi}}{d\tau}=&\Re\left\{\sum_{n,m}\mathbf{G}A_{n,m}\left[e^{i\left[\mathbf{G}\cdot\left(\mathbf{P}+\boldsymbol{\eta}-\mathbf{g}\right)-\pi/2\right]}-\left\langle e^{i\left[\mathbf{G}\cdot\left(\mathbf{P}+\boldsymbol{\eta}-\mathbf{g}\right)-\pi/2\right]}\right\rangle\right]\right\},\label{eq:rnr}\\
\omega\frac{d\boldsymbol{\eta}}{d\tau}=&-\mathbf{\nabla}V(\mathbf{R}+\boldsymbol{\xi})+\left\langle\mathbf{\nabla}V(\mathbf{R}+\boldsymbol{\xi})\right\rangle,\label{eq:knr}
\end{align}
the system of equations \eqref{eq:rnr}, \eqref{eq:knr} constitutes the \textit{``effective equations of motion for the fast coordinates''}.

\subsection{Asymptotic expansion in powers of $\epsilon=1/\omega$}

Since $\omega\rightarrow\infty$ implies $\epsilon=1/\omega\rightarrow0$, we propose the expansions for the fast coordinates as:
\begin{align}
\boldsymbol{\xi}&=\sum^{\infty}_{i=0}\epsilon^{i}\boldsymbol{\xi}_i=\boldsymbol{\xi}_{0}+\epsilon\boldsymbol{\xi}_{1}+\epsilon^{2}\boldsymbol{\xi}_{2}+\cdots,\label{eq:crx}\\
\boldsymbol{\eta}&=\sum^{\infty}_{i=0}\epsilon^{i}\boldsymbol{\eta}_{i}=\boldsymbol{\eta}_{0}+\epsilon\boldsymbol{\eta}_{1}+\epsilon^{2}\boldsymbol{\eta}_{2}+\cdots,\label{eq:crp}
\end{align}
truncated at $O(\epsilon^2)$. Substituting the series \eqref{eq:crx} and \eqref{eq:crp} into equations \eqref{eq:rnr} and \eqref{eq:knr}, we define the auxiliary functions:
\begin{equation}
\alpha(\epsilon)\equiv e^{i[\mathbf G\cdot(\mathbf P+\boldsymbol\eta_0+\epsilon\boldsymbol\eta_1+\epsilon^2\boldsymbol\eta_2+\cdots-\mathbf g)-\pi/2]},\qquad
\boldsymbol\beta(\epsilon)\equiv\mathbf R+\boldsymbol\xi_0+\epsilon\boldsymbol\xi_1+\epsilon^2\boldsymbol\xi_2+\cdots,
\label{eq:alfabeta}
\end{equation}
so that:
\begin{equation}
\frac{\dif\boldsymbol\xi_0}{\dif\tau}+\epsilon\frac{\dif\boldsymbol\xi_1}{\dif\tau}+\epsilon^2\frac{\dif\boldsymbol\xi_2}{\dif\tau}+\cdots=\epsilon\,\Re\!\left[\sum_{n,m}\mathbf G A_{n,m}(\alpha-\langle\alpha\rangle)\right],
\label{eq:rnr1}
\end{equation}
\begin{equation}
\frac{\dif\boldsymbol\eta_0}{\dif\tau}+\epsilon\frac{\dif\boldsymbol\eta_1}{\dif\tau}+\epsilon^2\frac{\dif\boldsymbol\eta_2}{\dif\tau}+\cdots=-\epsilon\big[\nabla V(\boldsymbol\beta)-\langle\nabla V(\boldsymbol\beta)\rangle\big].
\label{eq:knr1}
\end{equation}
Writing $\alpha(\epsilon)=e^{i[\mathbf G\cdot(\mathbf P+\boldsymbol\eta_0-\mathbf g)-\pi/2]}\cdot e^{i\mathbf G\cdot(\epsilon\boldsymbol\eta_1+\epsilon^2\boldsymbol\eta_2+\cdots)}$ and expanding the exponential in the second factor by means of the Taylor series $e^{iu}=1+iu-u^2/2+O(u^3)$ with $u=\mathbf G\cdot(\epsilon\boldsymbol\eta_1+\epsilon^2\boldsymbol\eta_2)+O(\epsilon^3)=\epsilon(\mathbf G\cdot\boldsymbol\eta_1)+\epsilon^2(\mathbf G\cdot\boldsymbol\eta_2)+O(\epsilon^3)$, we obtain, keeping terms up to $\epsilon^2$,
\begin{align}
e^{iu}&=1+i\big[\epsilon(\mathbf G\cdot\boldsymbol\eta_1)+\epsilon^2(\mathbf G\cdot\boldsymbol\eta_2)\big]-\frac{1}{2}\big[\epsilon(\mathbf G\cdot\boldsymbol\eta_1)\big]^2+O(\epsilon^3)\nonumber\\
&=1+i(\mathbf G\cdot\boldsymbol\eta_1)\epsilon+\left[i(\mathbf G\cdot\boldsymbol\eta_2)-\frac{(\mathbf G\cdot\boldsymbol\eta_1)^2}{2}\right]\epsilon^2+O(\epsilon^3),
\end{align}
from which
\begin{equation}
\alpha(\epsilon)=e^{i[\mathbf G\cdot(\mathbf P+\boldsymbol\eta_0-\mathbf g)-\pi/2]}\left\{1+i(\mathbf G\cdot\boldsymbol\eta_1)\epsilon+\left[i(\mathbf G\cdot\boldsymbol\eta_2)-\frac{(\mathbf G\cdot\boldsymbol\eta_1)^2}{2}\right]\epsilon^2\right\}+O(\epsilon^3).
\label{eq:exptey1}
\end{equation}

On the other hand, from \eqref{eq:alfabeta} let $\boldsymbol\beta(\epsilon)=\boldsymbol\beta_0+\Delta\boldsymbol\beta$, where $\boldsymbol\beta_0=\mathbf R+\boldsymbol\xi_0$ and $\Delta\boldsymbol\beta=\epsilon\boldsymbol\xi_1+\epsilon^2\boldsymbol\xi_2+O(\epsilon^3)$. Expanding the vector field $\nabla V(\boldsymbol\beta(\epsilon))$ defined in \eqref{eq:knr1} component by component ($i=1,2$ for $x,y$) in a multivariable Taylor series around $\Delta\boldsymbol\beta\rightarrow\boldsymbol0$, equivalent to expanding around $\epsilon\rightarrow0$:
\begin{equation}
[\nabla V(\boldsymbol\beta)]_i=\partial_iV(\boldsymbol\beta_0)+\sum_j\partial_i\partial_jV(\boldsymbol\beta_0)\,\Delta\beta_j+\frac12\sum_{j,k}\partial_i\partial_j\partial_kV(\boldsymbol\beta_0)\,\Delta\beta_j\Delta\beta_k+O(\Delta\beta^3).
\label{eq:taylorV}
\end{equation}
Substituting $\Delta\beta_j=\epsilon(\xi_1)_j+\epsilon^2(\xi_2)_j+O(\epsilon^3)$ and keeping only terms up to $\epsilon^2$ (note that $\Delta\beta_j\Delta\beta_k=\epsilon^2(\xi_1)_j(\xi_1)_k+O(\epsilon^3)$, since the cross term $\xi_1\xi_2$ first appears at order $\epsilon^3$):
\begin{equation}
\begin{split}
[\nabla V(\boldsymbol\beta)]_i={}&\partial_iV(\boldsymbol\beta_0)+\epsilon\sum_jH_{ij}(\boldsymbol\beta_0)(\xi_1)_j\\
&+\epsilon^2\left[\frac12\sum_{j,k}T_{ijk}(\boldsymbol\beta_0)(\xi_1)_j(\xi_1)_k+\sum_jH_{ij}(\boldsymbol\beta_0)(\xi_2)_j\right]+O(\epsilon^3),
\end{split}
\label{eq:exptey2idx}
\end{equation}
where $H_{ij}\equiv\partial^2V/\partial x_i\partial x_j$ is the Hessian matrix of $V$ and $T_{ijk}\equiv\partial^3V/\partial x_i\partial x_j\partial x_k$ the third-derivative tensor. In compact vector-tensor notation,
\begin{equation}
\nabla V(\boldsymbol\beta(\epsilon))=\nabla V(\boldsymbol\beta_0)+\big[\mathbf H(\boldsymbol\beta_0)\cdot\boldsymbol\xi_1\big]\epsilon
+\left[\frac12\mathbf T(\boldsymbol\beta_0):(\boldsymbol\xi_1\otimes\boldsymbol\xi_1)+\mathbf H(\boldsymbol\beta_0)\cdot\boldsymbol\xi_2\right]\epsilon^2+O(\epsilon^3),
\label{eq:exptey2}
\end{equation}
where $[\mathbf T:(\boldsymbol\xi_1\otimes\boldsymbol\xi_1)]_i\equiv\sum_{j,k}T_{ijk}(\xi_1)_j(\xi_1)_k$ and $[\mathbf H\cdot\boldsymbol\xi]_i\equiv\sum_jH_{ij}\xi_j$ is the usual matrix-vector product.

\subsection{Solution order by order in powers of $\epsilon$}

Substituting equations \eqref{eq:exptey1} and \eqref{eq:exptey2} into the system of equations \eqref{eq:rnr1} and \eqref{eq:knr1}, and comparing the terms of the resulting equations order by order in powers of $\epsilon$, we obtain:

\emph{Order $\epsilon^0$:}
\begin{align}
\frac{\dif\boldsymbol\xi_0}{\dif\tau}=&0,\label{eq:xi0}\\
\frac{\dif\boldsymbol\eta_0}{\dif\tau}=&0.\label{eq:eta0}
\end{align}
Integrating equations \eqref{eq:xi0} and \eqref{eq:eta0} yields constants (independent of $\tau$); these are chosen equal to zero so that the functions $\boldsymbol{\xi}$ and $\boldsymbol{\eta}$ have zero time average, as required by \eqref{eq:corrap}: $\boldsymbol{\xi}_{0}=\boldsymbol{\eta}_{0}=0$.

\emph{Order $\epsilon^1$:}\\
Using $\boldsymbol\eta_0=\mathbf 0$ in \eqref{eq:exptey1}, the right-hand side of \eqref{eq:rnr1} at this order is
\[
\Re\Big\{\sum\mathbf GA_{n,m}e^{i[\mathbf G\cdot(\mathbf P-\mathbf g)-\pi/2]}\Big\}-\Re\Big\{\sum\mathbf GA_{n,m}\langle e^{i[\cdots]}\rangle\Big\},
\]
that is,
\begin{equation}
\frac{\dif\boldsymbol\xi_1}{\dif\tau}=\Re\left\{\sum_{n,m}\mathbf G A_{n,m}e^{i(\mathbf G\cdot\mathbf P-\pi/2)}\Big[e^{-i\mathbf G\cdot\mathbf g}-\big\langle e^{-i\mathbf G\cdot\mathbf g}\big\rangle\Big]\right\},
\label{eq:xi1}
\end{equation}
and substituting $\boldsymbol\xi_0=\mathbf 0$ into \eqref{eq:exptey2}, the term corresponding to this order according to \eqref{eq:knr1}, evaluating \eqref{eq:exptey2} at $\boldsymbol\beta_0=\mathbf R$,
\begin{equation}
\frac{\dif\boldsymbol\eta_1}{\dif\tau}=-\big[\nabla V(\mathbf R)-\langle\nabla V(\mathbf R)\rangle\big]=\mathbf 0
\quad\Longrightarrow\quad \boldsymbol\eta_1=\mathbf 0,
\label{eq:eta1}
\end{equation}
where the vanishing in \eqref{eq:eta1} follows because $\mathbf R$ is a slow coordinate and $\langle\mathbf R\rangle=\mathbf R$ by \eqref{eq:corlen}, so that $\nabla V(\mathbf R)$ does not depend on $\tau$ and coincides with its own average; hence integrating with the zero-average condition gives $\boldsymbol\eta_1=\mathbf 0$.

\emph{Order $\epsilon^2$:}\\
The term $\mathbf G\cdot\boldsymbol\eta_1$ appearing in \eqref{eq:exptey1} vanishes by \eqref{eq:eta1}, and the right-hand side of \eqref{eq:rnr1} has no new contribution at this order, so
\begin{equation}
\frac{\dif\boldsymbol\xi_2}{\dif\tau}=\mathbf 0\ \Longrightarrow\ \boldsymbol\xi_2=\mathbf 0,
\label{eq:xi2}
\end{equation}
from \eqref{eq:exptey2} there is a contribution at this order due to the term $\mathbf H(\mathbf R)\cdot\boldsymbol\xi_1$ with $\boldsymbol\xi_1\neq\mathbf 0$ as given by \eqref{eq:xi1}, so:
\begin{equation}
\frac{\dif\boldsymbol\eta_2}{\dif\tau}=-\mathbf H(\mathbf R)\cdot\boldsymbol\xi_1.
\label{eq:eta2}
\end{equation}

With these results, the effective equations of motion for $\mathbf{R}(t)$ and $\mathbf{P}(t)$ given in \eqref{eq:rnn1} and \eqref{eq:knn1} become
\begin{align}
\dot{\mathbf R}&=\Re\left\{\sum_{n,m}\mathbf G A_{n,m}\left[\Big\langle e^{i[\mathbf G\cdot(\mathbf P-\mathbf g)-\pi/2]}\Big\rangle+\epsilon^2\Big\langle i(\mathbf G\cdot\boldsymbol\eta_2)\,e^{i[\mathbf G\cdot(\mathbf P-\mathbf g)-\pi/2]}\Big\rangle\right]\right\},\label{eq:ref1}\\
\dot{\mathbf P}&=-\nabla V(\mathbf R)-\frac{\epsilon^2}{2}\,\mathbf T(\mathbf R):\big\langle\boldsymbol\xi_1\otimes\boldsymbol\xi_1\big\rangle,\label{eq:kef1}
\end{align}
where \eqref{eq:kef1} is obtained by averaging \eqref{eq:exptey2} and using $\langle\boldsymbol\xi_1\rangle=0$ and $\boldsymbol\xi_2=\mathbf 0$.

\section{Fourier series expansion of $e^{-i\mathbf{G}\cdot\mathbf{g}(t)}$}

Since $\mathbf f(\omega t)$ is periodic, it admits a Fourier series expansion
\begin{equation}
\mathbf f(\omega t)=\sum_{k\in\mathbb Z}\mathbf f_k\,e^{ik\omega t}.\label{eq:fourierf}
\end{equation}
Since the oscillating field $\mathbf f(\omega t)$ is a real physical field, its coefficients satisfy the standard condition
\begin{equation}
\mathbf f_{-k}=\overline{\mathbf f_k},
\label{eq:realf}
\end{equation}
where $\overline{\mathbf f_k}$ denotes the complex conjugate of $\mathbf f_{k}$. Recalling the definition $\mathbf{g}(t)\equiv\int^{t}_{0}\mathbf{f}(\omega t')dt'$ and integrating \eqref{eq:fourierf} term by term, separating the case $k=0$, which gives a term that grows without bound in time of the form $\mathbf f_0t$, from the cases $k\neq0$, for which we apply integration of the type $\int_0^te^{ik\omega t'}\dif t'=(e^{ik\omega t}-1)/(ik\omega)$, we obtain:
\begin{equation}
\mathbf g(t)=\mathbf f_0t+\sum_{k\neq0}\mathbf f_k\,\frac{e^{ik\omega t}-1}{ik\omega}.
\label{eq:g}
\end{equation}
Then
\begin{equation}
\mathbf G\cdot\mathbf g(t)=\mathbf G\cdot\mathbf f_0 t+\sum_{k\neq0}\mathbf{G} \cdot \mathbf{f}_k\frac{e^{ik\omega t}-1}{ik\omega},
\label{eq:Gg}
\end{equation}
defining $\Omega\equiv\mathbf G\cdot\mathbf f_0$ and $\alpha_k\equiv\mathbf G\cdot\mathbf f_k/(ik\omega)$ for $k\neq0$, equation \eqref{eq:Gg} becomes:
\begin{equation}
\mathbf G\cdot\mathbf g(t)=\Omega t+\sum_{k\neq0}\alpha_k\big(e^{ik\omega t}-1\big).
\label{eq:Gg1}
\end{equation}

Furthermore, from the definition $\alpha_k=\mathbf G\cdot\mathbf f_k/(ik\omega)$ and using property \eqref{eq:realf} we have: $\alpha_{-k}=\mathbf G\cdot\mathbf f_{-k}/(-ik\omega)=\mathbf G\cdot\overline{\mathbf f_k}/(-ik\omega)$. On the other hand, the complex conjugate of $\alpha_k$ becomes $\overline{\alpha_k}=\overline{\mathbf G\cdot\mathbf f_k}/\overline{ik\omega}=\mathbf G\cdot\overline{\mathbf f_k}/(-ik\omega)$, since $\mathbf G,k,\text{and}\;\omega$ are real. Thus, we may conclude from this algebra that: $\alpha_{-k}=\overline{\alpha_k}$.

On the other hand, using \eqref{eq:Gg1}
\begin{equation}
e^{-i\mathbf{G}\cdot\mathbf{g}(t)} = e^{-i\Omega t}\exp{\left( -i \sum_{k \neq 0} \alpha_k (e^{i k \omega t} - 1) \right)},\label{eq:eGg}
\end{equation}
note that the term $\exp{\left( i \sum_{k \neq 0} \alpha_k \right)}$ does not depend on $t$, so we define the factor $C=C(n,m)=\exp\left( i \sum_{k \neq 0} \alpha_k \right) = \exp{\left( \sum_{k \neq 0}\mathbf{G} \cdot \mathbf{f}_k/(k \omega) \right)}$. Separating the sum over $k$ in the exponent of $C(n,m)$, which runs over all integers except $k=0$, and using the previous result:
\begin{equation}
i\sum_{k\neq0}\alpha_k=i\sum_{k>0}(\alpha_k+\alpha_{-k})=i\sum_{k>0}(\alpha_k+\overline{\alpha_k})=i\sum_{k>0}2\Re(\alpha_k)=2i\sum_{k>0}\Re(\alpha_k),\label{eq:alpr}
\end{equation}
which is a pure imaginary number. Therefore the factor $C(n,m)=\exp(\text{pure imaginary})$ has modulus $1$.

Next we establish some properties of the functions defined above under a sign change in $n$ and $m$: if $(n,m)$ is simultaneously changed to $(-n,-m)$, then $\mathbf G\to-\mathbf G$ and therefore $\alpha_k\to-\alpha_k$ for every value of $k$, so that $C(-n,-m)=\exp(-i\sum_{k\neq0}\alpha_k)=\overline{C(n,m)}$. Recalling from complex analysis that $C\overline{C}=|C|^2=1$, we conclude that the complex conjugate of $C$ coincides with its reciprocal.

With $C(n,m)$ defined above, equation \eqref{eq:eGg} becomes:
\begin{equation}
e^{-i\mathbf{G}\cdot\mathbf{g}(t)} = C \, e^{-i\Omega t} \, \exp\left( -i \sum_{k \neq 0} \alpha_k e^{i k \omega t} \right),\label{eq:eGg1}
\end{equation}
defining the function \(F(\tau) = \exp\left( -i \sum_{k \neq 0} \alpha_k e^{i k \tau} \right)\) appearing in \eqref{eq:eGg1}, note that $F(\tau)$ is periodic, so we expand it in a Fourier series:
\begin{equation}
F(\tau) = \sum_{\ell \in \mathbb{Z}} \beta_\ell \, e^{i \ell \tau},\label{eq:Fp}
\end{equation}
where \(\beta_\ell\) are the coefficients of the complex Fourier series expansion \eqref{eq:Fp}, which depend on \(\alpha_k\) as:
\begin{equation}
\beta_\ell =\beta_\ell(n,m)=\frac{1}{2\pi} \int_0^{2\pi} \exp\left( -i \sum_{k \neq 0} \alpha_k e^{i k \tau} - i \ell \tau \right) \dif\tau,\label{eq:beta}
\end{equation}
where $\tau=\omega t$; substituting \eqref{eq:Fp} into \eqref{eq:eGg1}
\begin{equation}
e^{-i\mathbf G\cdot\mathbf g(t)}=\sum_{\ell\in\mathbb Z}C\,\beta_\ell\,e^{i(\ell\omega-\Omega)t}.
\label{eq:expGg}
\end{equation}

Observe that under the simultaneous sign change in $n$ and $m$, the coefficients \(\beta_\ell\) satisfy a property analogous to that of the factor $C$. If the change of variables $(n,m)\to(-n,-m)$ is performed, then $\alpha_k\to-\alpha_k$ as noted above; on the other hand, the function \(F(\tau)\) defined above depends on \(\alpha_k\), which in turn depends on the integers $n$ and $m$ by its definition, so, strictly, it should be written in the form $F(\tau;n,m)=\exp\big(-i\sum_{k\neq0}\alpha_ke^{ik\tau}\big)$; then, with the change of variables $(n,m)\to(-n,-m)$, $F(\tau;-n,-m)=\exp\big(i\sum_{k\neq0}\alpha_ke^{ik\tau}\big)$. On the other hand, if the complex conjugate is applied to $F(\tau;n,m)$ and the previously proven property for \(\alpha_k\) is used:
\begin{equation}
\overline{F(\tau;n,m)}=\exp\Big(i\sum_{k\neq0}\overline{\alpha_k}e^{-ik\tau}\Big)
=\exp\Big(i\sum_{k\neq0}\alpha_{-k}e^{-ik\tau}\Big)
=\exp\Big(i\sum_{k\neq0}\alpha_ke^{ik\tau}\Big).\label{eq:conF}
\end{equation}
in the last step we relabeled $k\to-k$ in the sum, which is valid since the sum runs over all nonzero integers. Hence we may conclude preliminarily that: $F(\tau;-n,-m)=\overline{F(\tau;n,m)}$. Writing the Fourier expansion \eqref{eq:Fp} in the form $F(\tau;n,m)=\sum_\ell\beta_\ell(n,m)e^{i\ell\tau}$, its complex conjugate is $\overline{F(\tau;n,m)}=\sum_\ell\overline{\beta_\ell(n,m)}e^{-i\ell\tau}=\sum_\ell\overline{\beta_{-\ell}(n,m)}e^{i\ell\tau}$, where in the last step we relabeled $\ell\to-\ell$; comparing the above expansion with $F(\tau;-n,-m)=\sum_\ell\beta_\ell(-n,-m)e^{i\ell\tau}$ we conclude that: $\beta_\ell(-n,-m)=\overline{\beta_{-\ell}(n,m)}$.

In the case relevant for the rest of this work, we will take $\mathbf f_0=\mathbf 0$, that is, an oscillating field with zero time average, so that $\Omega=0$ and equation \eqref{eq:expGg} becomes:
\begin{equation}
e^{-i\mathbf G\cdot\mathbf g(t)}=\sum_{\ell\in\mathbb Z}C\,\beta_\ell\,e^{i\ell\omega t}=\sum_{\ell\in\mathbb Z}C\,\beta_\ell\,e^{i\ell\tau}.
\label{eq:expGg2}
\end{equation}

\section{Effective equations of motion for the time derivatives of the coordinates: position $\dot{\mathbf{R}}(t)$ and displaced momentum $\dot{\mathbf{P}}(t)$ up to order $\epsilon^2$}

We seek to integrate the effective equations of motion for the time derivatives of the position coordinate $\mathbf{R}(t)$ and the displaced momentum $\mathbf{P}(t)$ given in \eqref{eq:ref1} and \eqref{eq:kef1}, in order to find an effective Hamiltonian $H(\mathbf{R},\mathbf{P})$.
To begin, we take the time average of \eqref{eq:expGg2}
\begin{align}
\left\langle e^{-i\mathbf{G}\cdot\mathbf{g}(t)}\right\rangle=\sum_{\ell \in \mathbb{Z}} C\, \beta_\ell \,\left\langle e^{i \ell \tau}\right\rangle,\label{eq:prome1}
\end{align}
and using the orthogonality relation of complex exponentials over a period
\begin{align}
\left\langle e^{i(n\pm m)\tau}\right\rangle&=\delta_{n,\mp m},\label{eq:prome2}
\end{align}
then \eqref{eq:prome1} becomes:
\begin{align}
\left\langle e^{-i\mathbf{G}\cdot\mathbf{g}(t)}\right\rangle=C \, \beta_0.\label{eq:prome3}
\end{align}

\subsection{Integration of the equations for $\boldsymbol\xi_1$ and $\boldsymbol\eta_2$}
Substituting equation \eqref{eq:expGg2} and \eqref{eq:prome3} into \eqref{eq:xi1},
\begin{equation}
\frac{\dif\boldsymbol\xi_1}{\dif\tau}=\Re\left\{\sum_{n,m,\ell\neq0}\mathbf GA_{n,m}C\beta_\ell\,e^{i(\mathbf G\cdot\mathbf P-\pi/2)}e^{i\ell\tau}\right\},
\label{eq:etan1}
\end{equation}
note that the term corresponding to $\ell=0$ in \eqref{eq:xi1} cancels exactly against $\langle e^{-i\mathbf G\cdot\mathbf g}\rangle=C\beta_0$ from \eqref{eq:prome3}. Integrating term by term with respect to $\tau$, each summand is a pure exponential of the type $e^{i\ell\tau}$ whose antiderivative is $e^{i\ell\tau}/(i\ell)=-ie^{i\ell\tau}/\ell=e^{i(\ell\tau-\pi/2)}/\ell$, where we used the complex-number properties $1/i=-i=e^{-i\pi/2}$, and applying $e^{i(\mathbf G\cdot\mathbf P-\pi/2)}\cdot e^{-i\pi/2}=e^{i(\mathbf G\cdot\mathbf P-\pi)}$, we obtain as a result:
\begin{equation}
\boldsymbol\xi_1(\tau)=\Re\left\{\sum_{n,m,\ell\neq0}\mathbf GA_{n,m}C\beta_\ell\,e^{i(\mathbf G\cdot\mathbf P-\pi)}\,\frac{e^{i\ell\tau}}{\ell}\right\},
\label{eq:etan2}
\end{equation}
with an integration constant that vanishes by the zero-average condition \eqref{eq:corrap}.

Substituting \eqref{eq:etan2} into \eqref{eq:eta2}, $\dif\boldsymbol\eta_2/\dif\tau=-\mathbf H\cdot\boldsymbol\xi_1$, and using the linearity of the matrix-vector product: $\mathbf H\cdot(\mathbf G A_{n,m}\cdots)=(\mathbf H\cdot\mathbf G)A_{n,m}\cdots$, together with $-e^{i(\mathbf G\cdot\mathbf P-\pi)}=e^{i\mathbf G\cdot\mathbf P}$, since $e^{-i\pi}=-1$,
\begin{equation}
\frac{\dif\boldsymbol\eta_2}{\dif\tau}=\Re\left\{\sum_{n,m,\ell\neq0}(\mathbf H\cdot\mathbf G)A_{n,m}C\beta_\ell\,e^{i\mathbf G\cdot\mathbf P}\,\frac{e^{i\ell\tau}}{\ell}\right\}.
\label{eq:etan3}
\end{equation}
Integrating again term by term in $\tau$, with
\[
\int\frac{e^{i\ell\tau}}{\ell}\,\dif\tau=\frac{e^{i\ell\tau}}{i\ell^2}=\frac{e^{i(\ell\tau-\pi/2)}}{\ell^2},\qquad e^{i\mathbf G\cdot\mathbf P}e^{-i\pi/2}=e^{i(\mathbf G\cdot\mathbf P-\pi/2)},
\]
and with zero integration constant by the same zero-average condition \eqref{eq:corrap},
\begin{equation}
\boldsymbol\eta_2(\tau)=\Re\left\{\sum_{n,m,\ell\neq0}(\mathbf H\cdot\mathbf G)A_{n,m}C\beta_\ell\,e^{i(\mathbf G\cdot\mathbf P-\pi/2)}\,\frac{e^{i\ell\tau}}{\ell^2}\right\}.
\label{eq:etan4}
\end{equation}

\subsection{Calculation of the terms of $\dot{\mathbf R}$}

Next we develop each term of equation \eqref{eq:ref1}. The $\epsilon^0$-order term is obtained directly with the help of \eqref{eq:prome3}:
\begin{equation}
\Big\langle e^{i[\mathbf G\cdot(\mathbf P-\mathbf g)-\pi/2]}\Big\rangle=e^{i(\mathbf G\cdot\mathbf P-\pi/2)}\big\langle e^{-i\mathbf G\cdot\mathbf g}\big\rangle=C\beta_0\,e^{i(\mathbf G\cdot\mathbf P-\pi/2)}.
\label{eq:val1}
\end{equation}

For the $\epsilon^2$-order term of \eqref{eq:ref1}, we write
\begin{equation}
\mathbf G\cdot\boldsymbol\eta_2=\Re\{\sum_{n',m',\ell'\neq0}[\mathbf G\cdot(\mathbf H\cdot\mathbf G')]A_{n',m'}C'\beta_{\ell'}e^{i(\mathbf G'\cdot\mathbf P-\pi/2)}e^{i\ell'\tau}/\ell'^2\},
\label{eq:Geta2}
\end{equation}
where we use primed notation for the indices $(n',m',\ell')$ coming from $\boldsymbol\eta_2$, and we identify the triple scalar-matrix product as follows:\\
Let the vector be: $\mathbf{G}'=\begin{pmatrix} G'_1 \\ G'_2 \end{pmatrix}$ and the matrix: $\mathbf{H} = \begin{pmatrix} H_{11} & H_{12} \\ H_{21} & H_{22} \end{pmatrix}$. Then, the matrix-vector product
$$\mathbf{H}\cdot\mathbf{G}' = \begin{pmatrix} H_{11} & H_{12} \\ H_{21} & H_{22} \end{pmatrix} \begin{pmatrix} G'_1 \\ G'_2 \end{pmatrix} = \begin{pmatrix} H_{11}G'_1 + H_{12}G'_2 \\ H_{21}G'_1 + H_{22}G'_2 \end{pmatrix}.$$
The result is a new vector in $\mathbb{R}^2$, which we may call $\mathbf{w}$. We now take the dot product between $\mathbf{G}$ (row) and $\mathbf{w}$ (column):
$$\mathbf{G}\cdot\mathbf{w} = G_1 w_1 + G_2 w_2.$$
Substituting the components of $\mathbf{w}$:
\begin{equation}
\mathbf G\cdot(\mathbf H\cdot\mathbf G')=G_1(H_{11}G_1'+H_{12}G_2')+G_2(H_{21}G_1'+H_{22}G_2'),
\label{eq:GTHG}
\end{equation}
using the notation $\mathbf G^{T}$ for the transpose of the column matrix $\mathbf G$, equation \eqref{eq:GTHG} can be written in compact matrix notation as:
\begin{equation}
\mathbf G\cdot(\mathbf H\cdot\mathbf G')=\mathbf G^{T}\mathbf H\,\mathbf G',
\label{eq:GTHG1}
\end{equation}
that is, the outer dot simply transposes $\mathbf G$ to form a row vector, so that $\mathbf G\cdot(\mathbf H\cdot\mathbf G')$ is the scalar $\mathbf G^{T}\mathbf H\mathbf G'=\sum_{i,j}G_iH_{ij}G_j'$.

Using further $e^{-i\mathbf G\cdot\mathbf g(t)}=\sum_{\ell}C\beta_\ell e^{i\ell\tau}$ from \eqref{eq:expGg2} and the identity $i=e^{i\pi/2}$, the product $i(\mathbf G\cdot\boldsymbol\eta_2)e^{i[\mathbf G\cdot(\mathbf P-\mathbf g)-\pi/2]}$ is written as
\begin{equation}
\Re\left\{\sum_{n',m',\ell'\neq0}\mathbf G^{T}\mathbf H\mathbf G'A'C'\beta_{\ell'}e^{i(\mathbf G'\cdot\mathbf P-\pi/2)}\frac{e^{i\ell'\tau}}{\ell'^2}\right\}e^{i\mathbf G\cdot\mathbf P}\sum_{\ell}C\beta_\ell e^{i\ell\tau},
\label{eq:val2mid}
\end{equation}
where the factor $i\cdot(-i)=1$ coming from $i\,e^{-i\pi/2}$ has already been absorbed. Writing $\Re\{W\}=(W+\overline W)/2$ with $W\equiv\sum_{n',m',\ell'\neq0}\mathbf G^{T}\mathbf H\mathbf G'A'C'\beta_{\ell'}e^{i(\mathbf G'\cdot\mathbf P-\pi/2)}e^{i\ell'\tau}/\ell'^2$, and averaging over $\tau$ the product of \eqref{eq:val2mid} with $\Re\{W\}$, using $\langle e^{i(\ell+\ell')\tau}\rangle=\delta_{\ell',-\ell}$ and $\langle e^{i(\ell-\ell')\tau}\rangle=\delta_{\ell,\ell'}$, we obtain
\begin{align}
\Big\langle i(\mathbf G\cdot\boldsymbol\eta_2)\,e^{i[\mathbf G\cdot(\mathbf P-\mathbf g)-\pi/2]}\Big\rangle
=&\ \frac12\sum_{n',m',\ell\neq0}\mathbf G^{T}\mathbf H\mathbf G'A'CC'\beta_\ell\beta_{-\ell}'\,\frac{e^{i[(\mathbf G+\mathbf G')\cdot\mathbf P-\pi/2]}}{\ell^2}\nonumber\\
&+\frac12\sum_{n',m',\ell\neq0}\mathbf G^{T}\mathbf H\mathbf G'A'C\overline{C'}\beta_\ell\overline{\beta_\ell'}\,\frac{e^{i[(\mathbf G-\mathbf G')\cdot\mathbf P+\pi/2]}}{\ell^2},
\label{eq:val3}
\end{align}
where: $A'\equiv A_{n',m'}$, $\beta_{-\ell}'\equiv\beta_{-\ell}(n',m')$, and $\beta_\ell'\equiv\beta_\ell(n',m')$. Substituting \eqref{eq:val1} and \eqref{eq:val3} into \eqref{eq:ref1}, and multiplying by the overall factor $\mathbf GA_{n,m}$ followed by summation over $n$ and $m$:
\begin{align}
\dot{\mathbf R}=&\ \Re\Big\{\sum_{n,m}\mathbf GA_{n,m}C\beta_0\,e^{i(\mathbf G\cdot\mathbf P-\pi/2)}\Big\}\nonumber\\
&+\frac{\epsilon^2}{2}\Re\Big\{\sum_{n,n',m,m',\ell\neq0}\mathbf G\big[\mathbf G^{T}\mathbf H\mathbf G'\big]AA'CC'\beta_\ell\beta_{-\ell}'\,\frac{e^{i[(\mathbf G+\mathbf G')\cdot\mathbf P-\pi/2]}}{\ell^2}\Big\}\nonumber\\
&+\frac{\epsilon^2}{2}\Re\Big\{\sum_{n,n',m,m',\ell\neq0}\mathbf G\big[\mathbf G^{T}\mathbf H\mathbf G'\big]AA'C\overline{C'}\beta_\ell\overline{\beta_\ell'}\,\frac{e^{i[(\mathbf G-\mathbf G')\cdot\mathbf P+\pi/2]}}{\ell^2}\Big\}.
\label{eq:ref1n}
\end{align}

\subsection{Calculation of the terms of $\dot{\mathbf P}$}

For the second term of \eqref{eq:kef1}, we first use the identity already justified in Section~3, since $\boldsymbol\xi_1$ does not depend on $\mathbf R$ but only on $\mathbf P$ and $\tau$, so that $\nabla_{\mathbf R}$ commutes with $\langle\cdot\rangle_\tau$ and acts only on $\mathbf H(\mathbf R)$, thus:
\begin{equation}
\mathbf T(\mathbf R):\big\langle\boldsymbol\xi_1\otimes\boldsymbol\xi_1\big\rangle=\big\langle\mathbf T(\mathbf R):\boldsymbol\xi_1\otimes\boldsymbol\xi_1\big\rangle
=\big\langle\nabla_{\mathbf R}\!\big(\boldsymbol\xi_1^{T}\mathbf H(\mathbf R)\boldsymbol\xi_1\big)\big\rangle=\nabla_{\mathbf R}\Big(\big\langle\boldsymbol\xi_1^{T}\mathbf H(\mathbf R)\boldsymbol\xi_1\big\rangle\Big).
\label{eq:Txi}
\end{equation}
To verify equation \eqref{eq:Txi}, in components,
\begin{align*}
[\mathbf T:(\boldsymbol\xi_1\otimes\boldsymbol\xi_1)]_i&=\sum_{j,k}T_{ijk}(\xi_1)_j(\xi_1)_k=\sum_{j,k}\partial_iH_{jk}(\xi_1)_j(\xi_1)_k\\
&=\partial_i\Big(\sum_{j,k}H_{jk}(\xi_1)_j(\xi_1)_k\Big)=\partial_i(\boldsymbol\xi_1^{T}\mathbf H\boldsymbol\xi_1),
\end{align*}
where we used $T_{ijk}=\partial_iH_{jk}$ and the fact that $(\xi_1)_j$ and $(\xi_1)_k$ do not depend on $\mathbf R$.

Using \eqref{eq:etan2}, we compute $\boldsymbol\xi_1^{T}\mathbf H\boldsymbol\xi_1=\Re\{X\}\Re\{Y\}$ with
\[
X=\sum_{n,m,\ell\neq0}\mathbf G^{T}A_{n,m}C\beta_\ell\, e^{i(\mathbf G\cdot\mathbf P-\pi)}\frac{e^{i\ell\tau}}{\ell},\qquad
Y=\sum_{n',m',\ell'\neq0}\mathbf H\mathbf G'A'C'\beta_{\ell'}\, e^{i(\mathbf G'\cdot\mathbf P-\pi)}\frac{e^{i\ell'\tau}}{\ell'}.
\]
Using the complex-algebra identity $\Re\{X\}\Re\{Y\}=\tfrac12\Re\{XY\}+\tfrac12\Re\{X\overline Y\}$, which follows from writing $\Re\{Z\}=(Z+\overline Z)/2$ for both factors and expanding the product $(X+\overline X)(Y+\overline Y)/4=\tfrac14(XY+X\overline Y+\overline XY+\overline X\overline Y)$, and noting that $\overline XY=\overline{X\overline Y}$, so that $\overline XY+X\overline Y=2\Re\{X\overline Y\}$ and $XY+\overline X\overline Y=2\Re\{XY\}$,
\begin{align}
\boldsymbol\xi_1^{T}\mathbf H\boldsymbol\xi_1=&\ \frac12\Re\Big\{\sum_{n,m,n',m',\ell\neq0,\ell'\neq0}\mathbf G^{T}\mathbf H\mathbf G'AA'CC'\beta_\ell\beta_{\ell'}\,e^{i[(\mathbf G+\mathbf G')\cdot\mathbf P-2\pi]}\frac{e^{i(\ell+\ell')\tau}}{\ell\ell'}\Big\}\nonumber\\
&+\frac12\Re\Big\{\sum_{n,m,n',m',\ell\neq0,\ell'\neq0}\mathbf G^{T}\mathbf H\mathbf G'AA'C\overline{C'}\beta_\ell\overline{\beta_{\ell'}}\,e^{i[(\mathbf G-\mathbf G')\cdot\mathbf P]}\frac{e^{i(\ell-\ell')\tau}}{\ell\ell'}\Big\},
\label{eq:val5}
\end{align}
using $e^{-2i\pi}=1$ and averaging \eqref{eq:val5} over $\tau$ with $\langle e^{i(\ell+\ell')\tau}\rangle=\delta_{\ell',-\ell}$, $\langle e^{i(\ell-\ell')\tau}\rangle=\delta_{\ell,\ell'}$,
\begin{align}
\big\langle\boldsymbol\xi_1^{T}\mathbf H\boldsymbol\xi_1\big\rangle=&\ \frac12\Re\Big\{-\sum_{n,m,n',m',\ell\neq0}\mathbf G^{T}\mathbf H\mathbf G'AA'CC'\beta_\ell\beta_{-\ell}'\,\frac{e^{i(\mathbf G+\mathbf G')\cdot\mathbf P}}{\ell^2}\Big\}\nonumber\\
&+\frac12\Re\Big\{\sum_{n,m,n',m',\ell\neq0}\mathbf G^{T}\mathbf H\mathbf G'AA'C\overline{C'}\beta_\ell\overline{\beta_\ell'}\,\frac{e^{i(\mathbf G-\mathbf G')\cdot\mathbf P}}{\ell^2}\Big\},
\label{eq:val6}
\end{align}
with \eqref{eq:Txi} and \eqref{eq:val6}, equation \eqref{eq:kef1} is written explicitly as
\begin{align}
\dot{\mathbf P}=-\nabla V(\mathbf R)-\frac{\epsilon^2}{4}\nabla_{\mathbf R}\Re\Big\{&-\sum_{n,m,n',m',\ell\neq0}\mathbf G^{T}\mathbf H(\mathbf R)\mathbf G'AA'CC'\beta_\ell\beta_{-\ell}'\,\frac{e^{i(\mathbf G+\mathbf G')\cdot\mathbf P}}{\ell^2}\nonumber\\
&+\sum_{n,m,n',m',\ell\neq0}\mathbf G^{T}\mathbf H(\mathbf R)\mathbf G'AA'C\overline{C'}\beta_\ell\overline{\beta_\ell'}\,\frac{e^{i(\mathbf G-\mathbf G')\cdot\mathbf P}}{\ell^2}\Big\}.
\label{eq:kef1n}
\end{align}

\section{Derivation of the effective Hamiltonian $H(\mathbf{R},\mathbf{P})$}

With the effective equations of motion for $\dot{\mathbf{R}}$ and $\dot{\mathbf{P}}$ given in \eqref{eq:ref1n} and \eqref{eq:kef1n} respectively, we will now compute the corresponding effective Hamiltonian $H(\mathbf{R},\mathbf{P})$; note that time no longer appears explicitly in this Hamiltonian, unlike in the initial Hamiltonian given in \eqref{eq:H}, as a consequence of the averaging process over the period $2\pi/\omega$.

The effective Hamiltonian $H_{\text{ef}}(\mathbf{R},\mathbf{P})\equiv H_{\text{ef}}$ must satisfy Hamilton's equations in the variables $\mathbf{R}$ and $\mathbf{P}$ according to:
\begin{align}
\dot{\mathbf{R}}&=\frac{\partial H_{\text{ef}}}{\partial\mathbf{P}},\;\;\;&\dot{\mathbf{P}}&=-\frac{\partial H_{\text{ef}}}{\partial\mathbf{R}}.\label{eq:ham1}
\end{align}

On the other hand, note that a dynamical system $(\dot{\mathbf R},\dot{\mathbf P})=(\dot{\mathbf R}(\mathbf R,\mathbf P),\dot{\mathbf P}(\mathbf R,\mathbf P))$ admits a Hamiltonian $H(\mathbf R,\mathbf P)$ such that $\dot{\mathbf R}=\partial H/\partial\mathbf P$, $\dot{\mathbf P}=-\partial H/\partial\mathbf R$ if and only if the differential form
\begin{equation}
\dif H=\partial H/\partial\mathbf P\cdot\dif\mathbf P+\partial H/\partial\mathbf R\cdot\dif\mathbf R=\dot{\mathbf R}\cdot\dif\mathbf P-\dot{\mathbf P}\cdot\dif\mathbf R,\label{eq:formadif}
\end{equation}
is exact.

Equation \eqref{eq:kef1} can be written, using the expression already justified in \eqref{eq:Txi}, as:
\begin{align}
\dot{\mathbf P}&=-\nabla_{\mathbf R} V(\mathbf R)-\frac{\epsilon^2}{2}\,\nabla_{\mathbf R}\Big(\big\langle\boldsymbol\xi_1^{T}\mathbf H(\mathbf R)\boldsymbol\xi_1\big\rangle\Big),\nonumber\\
&=-\nabla_{\mathbf R}\left(V(\mathbf R)+\frac{\epsilon^2}{2}\,\big\langle\boldsymbol\xi_1^{T}\mathbf H(\mathbf R)\boldsymbol\xi_1\big\rangle\right),\label{eq:kef12}
\end{align}
then, from the second equation of \eqref{eq:ham1}, $\dot{\mathbf{P}}=-\partial H_{\text{ef}}/\partial\mathbf{R}$, integrating in $\mathbf{R}$ gives:
\begin{align}
H_{\text{ef}}=&-\int\dot{\mathbf{P}}\cdot d\mathbf{R}+D(\mathbf{P}),\label{eq:ham2}
\end{align}
substituting \eqref{eq:kef12} into \eqref{eq:ham2}:
\begin{align}
H_{\text{ef}}&=V(\mathbf R)+\frac{\epsilon^2}{2}\,\big\langle\boldsymbol\xi_1^{T}\mathbf H(\mathbf R)\boldsymbol\xi_1\big\rangle+D(\mathbf{P}).\label{eq:ham3}
\end{align}

Before continuing with the purpose of this section, if we define the first term of the scalar function \eqref{eq:val6}:
\begin{equation}
T_1(n,m,n',m',\ell)\equiv\big[\mathbf G^{T}\mathbf H\mathbf G'\big]AA'CC'\beta_\ell\beta_{-\ell}'\,\frac{e^{i(\mathbf G+\mathbf G')\cdot\mathbf P}}{\ell^2}.\label{eq:primt}
\end{equation}
Relabeling the dummy summation indices $(n,m)\leftrightarrow(n',m')$, which does not change the value of the sum, we obtain:
\begin{equation}
T_1(n',m',n,m,\ell)=\big[\mathbf G'^{T}\mathbf H\mathbf G\big]A'AC'C\beta_\ell(n',m')\beta_{-\ell}(n,m)\,\frac{e^{i(\mathbf G'+\mathbf G)\cdot\mathbf P}}{\ell^2}.\label{eq:primt1}
\end{equation}
Using the property: $\left(\mathbf G^{T}\mathbf H\mathbf G'\right)^{T}=\mathbf G'^{T}\mathbf H^{T}\left(\mathbf G^{T}\right)^{T}=\mathbf G'^{T}\mathbf H\mathbf G$, since $\left(\mathbf G^{T}\right)^{T}=\mathbf G$ and $\mathbf H^{T}=\mathbf H$, because $\mathbf H$ is symmetric. Additionally relabeling $\ell\to-\ell$, we arrive at
\begin{equation}
\begin{split}
T_1(n',m',n,m,-\ell)&=\big[\mathbf G^{T}\mathbf H\mathbf G'\big]AA'CC'\beta_{-\ell}(n',m')\beta_\ell(n,m)\,\frac{e^{i(\mathbf G+\mathbf G')\cdot\mathbf P}}{\ell^2}\\
&=T_1(n,m,n',m',\ell).
\end{split}\label{eq:primt2}
\end{equation}
On the other hand, defining the multiple sum $S_1\equiv\sum_{n,m,n',m',\ell\neq0}\mathbf G\,T_1$, after the full relabeling, that is, interchange of primed and unprimed indices followed by $\ell\to-\ell$, it satisfies: $S_1=\sum_{n',m',n,m,-\ell\neq0}\mathbf G'\,T_1$ with the same summand $T_1$ unrelabeled in view of \eqref{eq:primt2}. Therefore,
\begin{equation}
\sum\mathbf G\,T_1=\sum\mathbf G'\,T_1\quad\Longrightarrow\quad \sum\mathbf G\,T_1=\frac12\sum(\mathbf G+\mathbf G')\,T_1,
\label{eq:simetriaT1}
\end{equation}
where we used the definition $\sum_{n,m,n',m',\ell\neq0}=\sum_{n,m,n',m',-\ell\neq0}\equiv\sum$. On the other hand, we define for the second term of the scalar function \eqref{eq:val6}:
\begin{equation}
T_2(n,m,n',m',\ell)\equiv\big[\mathbf G^{T}\mathbf H\mathbf G'\big]AA'C\overline{C'}\beta_\ell\overline{\beta_\ell'}\,\frac{e^{i(\mathbf G-\mathbf G')\cdot\mathbf P}}{\ell^2}.\label{eq:segt2}
\end{equation}
Relabeling $(n,m)\leftrightarrow(n',m')$ without changing $\ell$,
\begin{equation}
T_2(n',m',n,m,\ell)=\big[\mathbf G^{T}\mathbf H\mathbf G'\big]AA'\overline{C}C'\overline{\beta_\ell}\beta_\ell'\,\frac{e^{i(\mathbf G'-\mathbf G)\mathbf P}}{\ell^2}=\overline{T_2(n,m,n',m',\ell)},\label{eq:segt21}
\end{equation}
since
\[
\overline{C\overline{C'}}=\overline{C}C',\qquad \overline{\beta_\ell\overline{\beta_\ell'}}=\overline{\beta_\ell}\beta_\ell',\qquad \overline{e^{i(\mathbf G-\mathbf G')\cdot\mathbf P}}=e^{i(\mathbf G'-\mathbf G)\cdot\mathbf P}.
\]
Likewise, defining $U_2(n,m,n',m',\ell)\equiv i\,T_2$, and redefining $(n,m)\leftrightarrow(n',m')$ without varying $\ell$ in $U_2$ and using \eqref{eq:segt21},
\[
U_2(n',m',n,m,\ell)=i\,\overline{T_2}=-\overline{i\,T_2}=-\overline{U_2}.
\]
Defining the multiple sum $S_2\equiv\sum_{n,m,n',m',\ell\neq0}\mathbf G\,U_2$, relabeling the primed and unprimed indices while keeping $\ell$ fixed:
\begin{equation}
S_2=\sum\mathbf G'\,\big(-\overline{U_2}\big)=-\overline{\sum\mathbf G'U_2}\quad\rightarrow\quad
\Re\{S_2\}=-\Re\Big\{\sum\mathbf G'U_2\Big\},
\label{eq:simetriaT2}
\end{equation}
then:
\begin{equation}
\Re\Big\{\sum\mathbf GU_2\Big\}=\frac12\Re\Big\{\sum(\mathbf G-\mathbf G')U_2\Big\}.
\label{eq:simetriaT3}
\end{equation}

Rewriting equation \eqref{eq:ref1n}, further using the identity $e^{\mp i\pi/2}=\mp i$ and the definitions \eqref{eq:primt}, \eqref{eq:segt2}:
\begin{align}
\dot{\mathbf R}=&\ \Re\Big\{\sum_{n,m}\mathbf GA_{n,m}C\beta_0\,e^{i(\mathbf G\cdot\mathbf P-\pi/2)}\Big\}+\frac{\epsilon^2}{2}\Re\Big\{-i\sum\mathbf G\,T_1\Big\}+\frac{\epsilon^2}{2}\Re\Big\{\sum\mathbf G\,U_2\Big\}.
\label{eq:ref1n1}
\end{align}
Substituting \eqref{eq:simetriaT1} and \eqref{eq:simetriaT3} into \eqref{eq:ref1n1}:
\begin{align}
\dot{\mathbf R}=&\ \Re\Big\{\sum_{n,m}\mathbf GA_{n,m}C\beta_0\,e^{i(\mathbf G\cdot\mathbf P-\pi/2)}\Big\}\nonumber\\
&+\frac{\epsilon^2}{4}\Re\Big\{-i\sum(\mathbf G+\mathbf G')\,T_1\Big\}+\frac{\epsilon^2}{4}\Re\Big\{i\sum(\mathbf G-\mathbf G')\,T_2\Big\}.
\label{eq:ref1n2}
\end{align}

On the other hand, differentiating \eqref{eq:val6} directly with respect to $\mathbf P$, note that the factors $\mathbf G^{T}\mathbf H\mathbf G',A,A',C,C'$ and $\beta$ do not depend on $\mathbf P$; only the exponentials $e^{i(\mathbf G+\mathbf G')\cdot\mathbf P}$ and $e^{i(\mathbf G-\mathbf G')\cdot\mathbf P}$ depend on $\mathbf P$, whose derivatives with respect to $\mathbf P$ are $i(\mathbf G+\mathbf G')e^{i(\mathbf G+\mathbf G')\cdot\mathbf P}$ and $i(\mathbf G-\mathbf G')e^{i(\mathbf G-\mathbf G')\cdot\mathbf P}$ respectively:
\begin{equation}
\frac{\partial}{\partial\mathbf P}\left(\big\langle\boldsymbol\xi_1^{T}\mathbf H\boldsymbol\xi_1\big\rangle\right)=\frac{1}{2}\Re\Big\{-i\sum(\mathbf G+\mathbf G')\,T_1\Big\}+\frac12\Re\Big\{i\sum(\mathbf G-\mathbf G')\,T_2\Big\},
\label{eq:Phifinal}
\end{equation}
where definitions \eqref{eq:primt} and \eqref{eq:segt2} were used. Differentiating \eqref{eq:ham3} with respect to $\mathbf P$:
\begin{align}
\frac{\partial H_{\text{ef}}}{\partial\mathbf{P}}&=\frac{\epsilon^2}{2}\frac{\partial}{\partial\mathbf P}\left(\big\langle\boldsymbol\xi_1^{T}\mathbf H\boldsymbol\xi_1\big\rangle\right)+\frac{\partial D(\mathbf{P})}{\partial\mathbf{P}}.\label{eq:dham3}
\end{align}
From the first equation of \eqref{eq:ham1}, $\dot{\mathbf{R}}=\partial H_{\text{ef}}/\partial\mathbf{P}$, substituting \eqref{eq:ref1n2} and \eqref{eq:Phifinal} into \eqref{eq:dham3}:
\begin{align}
&\Re\Big\{\sum_{n,m}\mathbf GA_{n,m}C\beta_0\,e^{i(\mathbf G\cdot\mathbf P-\pi/2)}\Big\}+\frac{\epsilon^2}{4}\Re\Big\{-i\sum(\mathbf G+\mathbf G')\,T_1\Big\}+\frac{\epsilon^2}{4}\Re\Big\{i\sum(\mathbf G-\mathbf G')\,T_2\Big\}\nonumber\\
&=\frac{\epsilon^2}{4}\Re\Big\{-i\sum(\mathbf G+\mathbf G')\,T_1\Big\}+\frac{\epsilon^2}{4}\Re\Big\{i\sum(\mathbf G-\mathbf G')\,T_2\Big\}+\frac{\partial D(\mathbf{P})}{\partial\mathbf{P}},\label{eq:dham4}
\end{align}
whence, simplifying \eqref{eq:dham4} we get:
\begin{align}
\Re\Big\{\sum_{n,m}\mathbf GA_{n,m}C\beta_0\,e^{i(\mathbf G\cdot\mathbf P-\pi/2)}\Big\}=\frac{\partial D(\mathbf{P})}{\partial\mathbf{P}},\label{eq:dham5}
\end{align}
integrating \eqref{eq:dham5} with respect to $\mathbf P$ and using $e^{-i\pi/2}=-i$:
\begin{align}
D(\mathbf{P})=-\Re\Big\{\sum_{n,m}A_{n,m}C\beta_0\,e^{i\mathbf G\cdot\mathbf P}\Big\}+\text{const.},\label{eq:dham6}
\end{align}
where the additive constant is set to zero without loss of generality. Substituting \eqref{eq:dham6} into \eqref{eq:ham3}, the effective Hamiltonian up to order $\epsilon^2=1/\omega^2$ is
\begin{equation}
\boxed{\begin{aligned}
H_{\mathrm{ef}}(\mathbf R,\mathbf P)={}&-\Re\Big\{\sum_{n,m}A_{n,m}\,C(n,m)\,\beta_0(n,m)\,e^{i\mathbf G\cdot\mathbf P}\Big\}+V(\mathbf R)\\
&+\frac{1}{2\omega^2}\Big\langle\boldsymbol\xi_1^{T}(\tau;\mathbf P)\,\mathbf H(\mathbf R)\,\boldsymbol\xi_1(\tau;\mathbf P)\Big\rangle_\tau
\end{aligned}}
\label{eq:Heffinal}
\end{equation}
where $\boldsymbol\xi_1(\tau;\mathbf P)$ is given explicitly by \eqref{eq:etan2} and the second-order term can be written explicitly using \eqref{eq:val6} as:
\begin{align}
&\frac{1}{2}\Big\langle\boldsymbol\xi_1^{T}(\tau;\mathbf P)\,\mathbf H(\mathbf R)\,\boldsymbol\xi_1(\tau;\mathbf P)\Big\rangle_\tau=\nonumber\\
&\frac14\Re\Big\{-\sum_{n,m,n',m',\ell\neq0}\big[\mathbf G^{T}\mathbf H(\mathbf R)\mathbf G'\big]A_{n,m}A_{n',m'}\nonumber\\
&\qquad\times C(n,m)C(n',m')\,\beta_\ell(n,m)\beta_{-\ell}(n',m')\,\frac{e^{i(\mathbf G+\mathbf G')\cdot\mathbf P}}{\ell^2}\nonumber\\
&+\sum_{n,m,n',m',\ell\neq0}\big[\mathbf G^{T}\mathbf H(\mathbf R)\mathbf G'\big]A_{n,m}A_{n',m'}\nonumber\\
&\qquad\times C(n,m)\overline{C(n',m')}\,\beta_\ell(n,m)\overline{\beta_\ell(n',m')}\,\frac{e^{i(\mathbf G-\mathbf G')\cdot\mathbf P}}{\ell^2}\Big\}.
\label{eq:correccionexplicita}
\end{align}
The first two terms of \eqref{eq:Heffinal}, $\varepsilon_0(\mathbf P)+V(\mathbf R)$, are the natural generalization of the static Hamiltonian, with $\varepsilon_0(\mathbf P)\equiv-\Re\Big\{\sum_{n,m}A_{n,m}\,C(n,m)\,\beta_0(n,m)\,e^{i\mathbf G\cdot\mathbf P}\Big\}$ representing the energy band renormalized by the oscillating field $\mathbf f(\omega t)$, in which each initial hopping element $A_{n,m}$ acquires a readjustment factor $C(n,m)\beta_0(n,m)$, characteristic of the Bessel-$J_0$ renormalization that appears in a standard single-harmonic problem, as can be verified in~\cite{mamani2017}. The order-$1/\omega^2$ term is a correction due to the oscillating field that modifies the particle's fast motion and, upon averaging that fast motion, an effective slower dynamics emerges; note also that this term is quadratic in the fast oscillation amplitude $\boldsymbol\xi_1$, and proportional to the local curvature of the static potential, $\mathbf H(\mathbf R)$.

\section{Results and discussion}
This section collects the particular cases of the general effective Hamiltonian \eqref{eq:Heffinal} that are used to illustrate its content: the two-dimensional lattice with zero static potential, the band freezing and dynamic localization conditions, a special two-dimensional field, and a field with mixed even--odd symmetry.

\textbf{Consistency with the one-dimensional results.} For $b=0$, $m=0$ and a field along $x$, Eq.~\eqref{eq:Heffinal} reduces to the one-dimensional effective Hamiltonian, and in that limit it reproduces the known results. For nearest-neighbor hopping, $A_{\pm1,0}=A_1$, and a monochromatic field $\mathbf f(\omega t)=2f_1\cos(\omega t)\,\mathbf i$, one finds $C(\pm1,0)=1$ and $\beta_0(\pm1,0)=J_0(2\lambda)$ with $\lambda\equiv af_1/\omega$, so that $\varepsilon_0(P)=-2A_1J_0(2\lambda)\cos(aP)$, i.e., the Dunlap--Kenkre renormalization $A_1\to A_1J_0(2\lambda)$~\cite{dunlap1986}; the order-$1/\omega^2$ term becomes $\Delta H=2A_1^2a^2V''(X)\,\omega^{-2}\big[S(\lambda)-\tilde S(\lambda)\cos(2aP)\big]$, with $S(\lambda)=\sum_{\ell\ge1}J_\ell^2(2\lambda)/\ell^2$ and $\tilde S(\lambda)=\sum_{\ell\ge1}(-1)^\ell J_\ell^2(2\lambda)/\ell^2$, and both $\Delta H$ and the renormalization vanish as $\omega\to\infty$, recovering the static nearest-neighbor Hamiltonian. The analogous reduction for first- and second-neighbor hopping under a bichromatic field, in which the factor $J_0$ is replaced by a generalized two-argument Bessel function, was obtained and discussed in Ref.~\cite{mamani2017} (Mamani, Calcina-Nogales, and Sanjin\'es Castedo), and is therefore not repeated here. Equation~\eqref{eq:Heffinal} recovers those first- and second-neighbor results as a particular case and extends them to hopping of arbitrary range and, with $b\neq0$, to two dimensions.

\textbf{Role of the symmetry of the field.} The symmetry of $\mathbf f(\omega t)$ under $\tau\to-\tau$ determines whether the renormalized band acquires a phase. For an even field the Fourier coefficients $\mathbf f_k$ are real, so each $\alpha_k$ is purely imaginary with $\alpha_{-k}=-\alpha_k$; hence $C(n,m)=1$ and $\beta_\ell(n,m)\in\mathbb R$, and the band $\varepsilon_0=-2\sum A_{n,m}\beta_0(n,m)\cos(\mathbf G\cdot\mathbf P)$ changes only in amplitude (a sign change of $\beta_0$ is a phase $\pi$), with no band phase shift (see Sec.~\ref{sec:even2D}). For an odd field, $\mathbf f_{-k}=-\mathbf f_k$ and $\mathbf f_k$ is purely imaginary, so each $\alpha_k$ is real and even in $k$; then $C(n,m)=e^{i\sum_{k\neq0}\alpha_k}$ is a nontrivial phase and $\beta_0$ is complex, so that each term of the band takes the form $\cos(\mathbf G\cdot\mathbf P+\arg(C\beta_0))$, a Peierls-type phase shift of the band, as illustrated by the mixed even--odd field of Sec.~\ref{sec:mixed}, where $C(n,m)=e^{-2im\lambda_y}$.

\subsection{Two-dimensional lattice with interaction to all neighbors and zero static potential}

We particularize~\eqref{eq:Heffinal} for the case:
\begin{equation}
\mathbf f(\omega t)=f_x(\omega t)\mathbf i+f_y(\omega t)\mathbf j,\qquad V(x,y)=0,
\label{eq:caso2D}
\end{equation}
that is, a two-dimensional lattice with interaction to all neighbors $(n,m)\neq(0,0)$ in the absence of the static potential. From the definition $H_{ij}(\mathbf R)\equiv\partial^2V/\partial x_i\partial x_j$ according to~\eqref{eq:exptey2idx}, with $V(x,y)=0$ we have:
\begin{equation}
\mathbf H(\mathbf R)\equiv\mathbf 0\quad\Longrightarrow\quad \boldsymbol\xi_1^T\mathbf H(\mathbf R)\boldsymbol\xi_1=0\quad\Longrightarrow\quad \frac{1}{2\omega^2}\big\langle\boldsymbol\xi_1^T\mathbf H(\mathbf R)\boldsymbol\xi_1\big\rangle_\tau=0.\label{eq:DeltaHcero}
\end{equation}
This shows an exact vanishing of the curvature term. Together with $V(\mathbf R)=0$, only the first term on the right-hand side of~\eqref{eq:Heffinal} survives.
\begin{equation}
H_{\mathrm{ef}}(\mathbf P)=-\Re\Big\{\sum_{n,m}A_{n,m}\,C(n,m)\,\beta_0(n,m)\,e^{i\mathbf G\cdot\mathbf P}\Big\}.\label{eq:Heffinalmod}
\end{equation}

\subsubsection{Linear structure of $\alpha_k(n,m)$, $C(n,m)$, and $\beta_\ell(n,m)$}
Writing the definition of $\mathbf G_{n,m}$ in the form: $\mathbf G_{n,m}=n\mathbf G_{1,0}+m\mathbf G_{0,1}$, where $\mathbf G_{1,0}\equiv a\,\mathbf i$ and $\mathbf G_{0,1}\equiv b\,\mathbf j$ are defined, then analogously for $\alpha_k(n,m)=n\,\alpha_k(1,0)+m\,\alpha_k(0,1)$, where $\alpha_k(1,0)=af_{x,k}/(ik\omega)$ and $\alpha_k(0,1)=bf_{y,k}/(ik\omega)$.

Substituting this linearity into $C(n,m)=\exp\big(i\sum_{k\neq0}\alpha_k(n,m)\big)$ and $F(\tau;n,m)=\exp\big(-i\sum_{k\neq0}\alpha_k(n,m)e^{ik\tau}\big)$, and defining $C_x\equiv C(1,0)$, $C_y\equiv C(0,1)$, $F_x(\tau)\equiv F(\tau;1,0)$ and $F_y(\tau)\equiv F(\tau;0,1)$, the functions $C(n,m)$, $F(\tau;n,m)$, and $\beta_\ell(n,m)$ can be written more compactly as:
\begin{align}
C(n,m)&=\left(C_x\right)^n\left(C_y\right)^m,\label{eq:Cnew}\\
F(\tau;n,m)&=\left[F_x(\tau)\right]^n\left[F_y(\tau)\right]^m,\label{eq:Fnew}\\
\beta_\ell(n,m)&=\frac{1}{2\pi}\int_0^{2\pi}\left[F_x(\tau)\right]^n\left[F_y(\tau)\right]^me^{-i\ell\tau}\dif\tau.\label{eq:betanew}
\end{align}
With $\alpha_{-k}=\overline{\alpha_k}$ already proven in Section~5, one shows that $|F(\tau;n,m)|=1$ and $|C(n,m)|=1$ for every $(n,m)$.

\subsubsection{Reduction of the sum}
Let
\[
\mathcal D\equiv\{(0,m):m>0\}\cup\{(n,0):n>0\}
\]
be a representative of each pair $\{(n,m),(-n,-m)\}$. Using the reflection symmetry of the hopping elements, $A_{-n,-m}=A_{n,m}$, and the properties already proven in Section~5,
\[
\beta_0(-n,-m)=\overline{\beta_0(n,m)},\qquad C(-n,-m)=\overline{C(n,m)},
\]
we obtain a simplified expression for the effective Hamiltonian \eqref{eq:Heffinalmod}:
\begin{equation}
H_{\mathrm{ef}}(\mathbf P)=-2\sum_{(n,m)\in\mathcal D}A_{n,m}\,\Re\big\{C_x^nC_y^m\,\beta_0(n,m)\,e^{i(naP_x+mbP_y)}\big\}.
\label{eq:Hef2Dgeneral}
\end{equation}

\subsubsection{Case of even $f_x,f_y$}\label{sec:even2D}
In this case, for each component we have $f_x(-\omega t)=f_x(\omega t)$ and $f_y(-\omega t)=f_y(\omega t)$; relabeling $k\to-k$ and using the condition that the oscillating field $\mathbf f(\omega t)$ is real, we obtain $f_{x,k},\,f_{y,k}\in\mathbb R$. As a consequence: $\alpha_k(1,0)$ and $\alpha_k(0,1)$ are purely imaginary, with $\alpha_{-k}(1,0)=-\alpha_k(1,0)$ and $\alpha_{-k}(0,1)=-\alpha_k(0,1)$, so that
\begin{equation}
C_x=\exp\Big(i\sum_{k>0}[\alpha_k(1,0)+\alpha_{-k}(1,0)]\Big)=1,\qquad C_y=\exp\Big(i\sum_{k>0}[\alpha_k(0,1)+\alpha_{-k}(0,1)]\Big)=1,\label{eq:CxCyUno}
\end{equation}
thus, from \eqref{eq:Cnew} we have: $C(n,m)=1$ $\forall(n,m)$.

Writing $\alpha_k(1,0)=-ir_k^x$ and $\alpha_k(0,1)=-ir_k^y$, where $r_k^x\equiv af_{x,k}/(k\omega)\in\mathbb R$ and $r_k^y\equiv bf_{y,k}/(k\omega)\in\mathbb R$ are defined and are moreover odd in $k$, the exponent of $F_x=F_x(\tau)\equiv F(\tau;1,0)$ is $g_x(\tau)=\sum_{k\neq0}\alpha_{k}(1,0)e^{ik\tau}=\sum_{k>0}\left[\alpha_{k}(1,0)e^{ik\tau}+\alpha_{-k}(1,0)e^{-ik\tau}\right]=2\sum_{k>0}r_k^x\sin(k\tau)$, which is real and odd; analogously, for the exponent of $F_y=F_y(\tau)\equiv F(\tau;0,1)$ we have $g_y(\tau)=\sum_{k\neq0}\alpha_{k}(0,1)e^{ik\tau}=\sum_{k>0}\left[\alpha_{k}(0,1)e^{ik\tau}+\alpha_{-k}(0,1)e^{-ik\tau}\right]=2\sum_{k>0}r_k^y\sin(k\tau)$, which is also real and odd. Then $F(\tau;n,m)=\left[F_x(\tau)\right]^n\left[F_y(\tau)\right]^m=e^{-ih_{n,m}(\tau)}$, where: $h_{n,m}(\tau)=ng_x(\tau)+mg_y(\tau)$ is real and odd, from which we see that: $F(-\tau;n,m)=\overline{F(\tau;n,m)}$, and, comparing the Fourier series expansions for both sides as in \eqref{eq:Fp}, we can conclude that:
\begin{equation}
\beta_\ell(n,m)\in\mathbb R\qquad\forall\,n,m\,\text{and}\,\ell.\label{eq:betaReal2D}
\end{equation}
Therefore, for a real oscillating field $\mathbf{f}(\omega t)$ with even symmetry, the effective Hamiltonian is obtained by substituting results \eqref{eq:CxCyUno} and \eqref{eq:betaReal2D} into the general expression \eqref{eq:Hef2Dgeneral}, giving:
\begin{equation}
H_{\mathrm{ef}}(\mathbf P)=-2\sum_{(n,m)\in\mathcal D}A_{n,m}\,\beta_0(n,m)\,\cos\big(naP_x+mbP_y\big).
\label{eq:Hef2DparFinal}
\end{equation}

\textbf{Limiting case:} with $b=0,m=0$, $\mathcal D\to\{n>0\}$, and~\eqref{eq:Hef2DparFinal} reduces to the one-dimensional even-field band $\varepsilon_0(P)=-2\sum_{n\geq1}A_n\beta_0(n,0)\cos(naP)$.

\subsubsection{Case of monochromatic fields $f_x=2f_1\cos\omega t$, $f_y=2f_2\cos\omega t$}

In this case, of all the terms for $f_{x,k}$ and $f_{y,k}$, only the following survive: $f_{x,\pm1}=f_1$ and $f_{y,\pm1}=f_2$, the rest vanish. Defining $\lambda_x\equiv af_1/\omega$, $\lambda_y\equiv bf_2/\omega$, and $\Lambda(n,m)\equiv n\lambda_x+m\lambda_y$:
\begin{equation}
\alpha_1(n,m)=-i\Lambda(n,m),\qquad\alpha_{-1}(n,m)=i\Lambda(n,m),\qquad\alpha_k(n,m)=0\qquad\forall|k|\neq1.\label{eq:alfaCosCos}
\end{equation}
On the other hand, the exponent argument of the function $F(\tau;n,m)$ for this case will be:
\begin{equation}
\begin{split}
\sum_{k\neq0}\alpha_k(n,m)e^{ik\tau}&=-i\Lambda(n,m)(e^{i\tau}-e^{-i\tau})=2\Lambda(n,m)\sin\tau\\
&\Longrightarrow\quad F(\tau;n,m)=e^{-2i\Lambda(n,m)\sin\tau}.
\end{split}\label{eq:FCosCos}
\end{equation}
Recalling the Jacobi--Anger relation: $e^{iz\sin\tau}=\sum_\ell J_\ell(z)e^{i\ell\tau}$, for this case with $z=-2\Lambda(n,m)$, and using the \emph{argument}-parity property of the Bessel function: $J_\ell(-x)=(-1)^\ell J_\ell(x)$, we have:
\begin{equation}
\beta_\ell(n,m)=(-1)^\ell J_\ell\big(2\Lambda(n,m)\big).
\label{eq:betaCosCos}
\end{equation}
With $J_{-\ell}(x)=(-1)^\ell J_\ell(x)$, which is the \emph{order}-parity property of the Bessel function, distinct from the previous property:
\begin{equation}
\beta_{-\ell}(n,m)=(-1)^{-\ell}J_{-\ell}(2\Lambda(n,m))=J_\ell\big(2\Lambda(n,m)\big).
\label{eq:betamenosellCosCos}
\end{equation}
In particular for this case: $\beta_0(n,m)=J_0(2\Lambda(n,m))\in\mathbb R$; and, using the property $\Lambda(-n,-m)=-\Lambda(n,m)$ (linearity) together with $J_0(-x)=J_0(x)$, we have:
\begin{equation}
\beta_0(-n,-m)=J_0(-2\Lambda(n,m))=J_0(2\Lambda(n,m))=\beta_0(n,m),
\label{eq:beta0simCosCos}
\end{equation}
which is consistent with the general property already proven in Section~5: $\beta_0(-n,-m)=\overline{\beta_0(n,m)}$. Finally, substituting into \eqref{eq:Hef2DparFinal}, we obtain the following effective Hamiltonian:
\begin{equation}
H_{\mathrm{ef}}(\mathbf P)=-2\sum_{(n,m)\in\mathcal D}A_{n,m}\,J_0\!\left(\frac{2(naf_1+mbf_2)}{\omega}\right)\cos(naP_x+mbP_y).
\label{eq:HefCosCosFinal}
\end{equation}

\textbf{Limiting case:} with $f_2=0$, $\Lambda(n,m)\to n\lambda_x$ and $\mathcal D\to\{n>0\}$,
\[
H_{\mathrm{ef}}\to-2\sum_{n\geq1}A_nJ_0(2n\lambda_x)\cos(naP_x),
\]
which is the one-dimensional single-harmonic renormalization $A_n\to A_nJ_0(2n\lambda_x)$ of Dunlap and Kenkre~\cite{dunlap1986}.

\subsection{Band freezing and dynamic localization}

In this section we analyze the particular case of the effective Hamiltonian \eqref{eq:Heffinal} obtained for $V(x,y)=0$ and a real oscillating field $\mathbf f(\omega t)$ with even symmetry; this case was already established in the preceding section, and we establish its precise relation to the exact dynamic localization (EDL) condition of the articles of Dignam \emph{et al.}~\cite{dignam2002} and Domachuk \emph{et al.}~\cite{domachuk2002}.

\subsubsection{Starting point}
For $V=0$ and $\mathbf f(\omega t)$ even, it was shown that
\begin{equation}
H_{\mathrm{ef}}(\mathbf P)=-2\sum_{(n,m)\in\mathcal D}A_{n,m}\,\beta_0(n,m)\,\cos(naP_x+mbP_y),
\label{eq:HefLocDin}
\end{equation}
with $C(n,m)=1$ and $\beta_0(n,m)\in\mathbb R$ for every $(n,m)$. From the general definition, eq.~\eqref{eq:prome3}, particularized for this case:
\begin{equation}
C(n,m)\,\beta_0(n,m)=\big\langle e^{-i\mathbf G_{n,m}\cdot\mathbf g(t)}\big\rangle=\frac{1}{T}\int^T_0 e^{-i\mathbf G_{n,m}\cdot\mathbf g(t)} dt,
\label{eq:defbeta0}
\end{equation}
an \emph{exact} quantity within the semiclassical formalism, obtained directly from the time average.

\subsubsection{Generalization of $S_p$ to two dimensions}
Starting from~\eqref{eq:defbeta0}, we define
\begin{equation}
S_{n,m}\equiv\big\langle e^{-i\mathbf G_{n,m}\cdot\mathbf g(t)}\big\rangle_t=C(n,m)\,\beta_0(n,m),
\label{eq:Snm}
\end{equation}
which is the direct generalization of Dignam's $S_p$~\cite{dignam2002}, to which it reduces exactly for $m=0$ and under the identification $n=p$, $a=d$, $f_x=E$, $e=\hbar=1$ already established. Under the parity hypothesis for $\mathbf f(\omega t)$, $C(n,m)=1$, and therefore
\begin{equation}
S_{n,m}=\beta_0(n,m),\label{eq:SnmIgualBeta}
\end{equation}
which is an exact, not approximate, identification. Note that the implication $\beta_0(n,m)=0\Rightarrow S_{n,m}=0$ holds \emph{unconditionally}, without needing the parity hypothesis on $\mathbf f(\omega t)$, since $|C(n,m)|=1$ is finite.

\subsubsection{Effective hopping elements and band freezing}
It is important to distinguish the initial parameters (hopping elements) $A_{n,m}$ of the two-dimensional Hamiltonian of the original lattice~\eqref{eq:H}, which are independent of the oscillating field $\mathbf f(\omega t)$, from the \emph{effective} hopping elements that arise from the effect of the rapidly oscillating field $\mathbf f(\omega t)$ and are defined by:
\begin{equation}
A_{n,m}^{\mathrm{ef}}\equiv A_{n,m}\,C(n,m)\,\beta_0(n,m)=A_{n,m}\,S_{n,m},
\label{eq:Aefectivo}
\end{equation}
which is the quantity that appears in~\eqref{eq:HefLocDin} for even $\mathbf f(\omega t)$. The condition
\begin{equation}
\beta_0(n,m)=0,\qquad\forall(n,m)\neq(0,0), \label{eq:condicionCentral}
\end{equation}
implies that $A_{n,m}^{\mathrm{ef}}=0\ \forall(n,m)\neq(0,0)$, but not that $A_{n,m}=0$, and by direct substitution into~\eqref{eq:HefLocDin}:
\begin{equation}
H_{\mathrm{ef}}(\mathbf P)\equiv0\qquad\forall\mathbf P. \label{eq:BandaPlana}
\end{equation}
Since the entire Fourier series of $H_{\mathrm{ef}}(\mathbf P)$ vanishes, in this context we will take \emph{flat band}, meaning $H_{\mathrm{ef}}=\mathrm{const.}=0$, and \emph{zero group velocity for every $\mathbf P$}, meaning $\nabla_{\mathbf P}H_{\mathrm{ef}}=0$, to be equivalent concepts.

\subsubsection{Dynamical consequence of band freezing}
From Hamilton's equations~\eqref{eq:ham1} applied to~\eqref{eq:BandaPlana}:
\begin{equation}
\dot{\mathbf R}=\frac{\partial H_{\mathrm{ef}}}{\partial\mathbf P}=0,\qquad
\dot{\mathbf P}=-\frac{\partial H_{\mathrm{ef}}}{\partial\mathbf R}=0
\quad\Longrightarrow\quad
\mathbf R(t)=\mathbf R(0),\ \ \mathbf P(t)=\mathbf P(0)\ \ \forall t,
\label{eq:congelamiento}
\end{equation}
which is an \emph{exact} result within the semiclassical framework (no additional hypotheses are required). From equation \eqref{eq:rs} we have $\mathbf r(t)=\mathbf R(t)+\boldsymbol\xi(\tau)$, which defines the position of the particle at an arbitrary instant $t$, split into a slow part and a fast part, and using the periodicity of $\boldsymbol\xi(\tau)$ in $\tau=\omega t$ with period $2\pi$:
\begin{equation}
\mathbf r(t+T)=\mathbf R(0)+\boldsymbol\xi(\tau+2\pi)=\mathbf R(0)+\boldsymbol\xi(\tau)=\mathbf r(t),\qquad T\equiv\frac{2\pi}\omega.
\label{eq:retornoPeriodico}
\end{equation}
This constitutes \textbf{dynamic localization in the semiclassical sense}: understood as the exact return of the position at every period $T$, with micromotions $\boldsymbol\xi(\tau)$ occurring between periods. This should not be confused with a static freezing at every instant $t$.

\subsubsection{Correspondence with the quantum formalism}

A connection can be established with the Floquet formulation. For a periodic quantum Hamiltonian \(\hat H(t+T)=\hat H(t)\), the evolution operator over one period defines the Floquet Hamiltonian via \cite{blanes2009}\cite{kuwahara2016}
\begin{equation}
U(T)= \mathcal{T}\exp\!\left[ -\frac{i}{\hbar}\int_0^T\hat H(t)\,dt \right] = \exp\!\left(-\frac{i}{\hbar}\hat H_FT\right),
\label{eq:operaun}
\end{equation}
where $\mathcal{T}$ is the time-ordering operator. In the high-frequency regime, $\hat{H}_F$ can be obtained by means of a Floquet--Magnus expansion \cite{bukov2015}. In \cite{rahav2003a} it has been shown that, under a consistent construction, the classical limit of this effective quantum Hamiltonian coincides with the classical effective Hamiltonian obtained through the separation of fast and slow scales. Consequently,
\begin{equation}
U(T)\simeq\exp\!\left(-\frac{i}{\hbar}H_{\mathrm{ef}}\,T\right),
\label{eq:FloquetCorrespondencia}
\end{equation}
the identification of the semiclassical effective Hamiltonian obtained here with a quantum Floquet Hamiltonian should be understood as an additional correspondence requiring the explicit establishment of the classical/quantum limit and the regime of validity of the expansion, rather than as an automatic consequence of the semiclassical derivation. Approximation \eqref{eq:FloquetCorrespondencia} is an additional external hypothesis — the Floquet--Magnus correspondence between the semiclassical effective Hamiltonian obtained here and a quantum Floquet Hamiltonian — which is standard in the literature \cite{mori2018}\cite{dey2025} but is not derived in this work. Under the additional hypothesis \eqref{eq:FloquetCorrespondencia}, if $H_{\mathrm{ef}}=0$ one would get $U(T)=I$, i.e., a trivial stroboscopic evolution, and hence $|\psi(nT)\rangle=|\psi(0)\rangle$ exactly every period —consistent with the definition of EDL of Dignam \emph{et al.}~\cite{dignam2002} and Domachuk \emph{et al.}~\cite{domachuk2002}—, but it \textbf{should not} be claimed that $|\psi(t)\rangle=|\psi(0)\rangle$ for arbitrary $t$, since that would ignore the intermediate micromotion evidenced in~\eqref{eq:retornoPeriodico}.

\subsection{Special band-freezing case for a type of two-dimensional external oscillating field}

We start from equation \eqref{eq:Heffinal}:
\begin{equation}
\begin{split}
H_{\mathrm{ef}}(\mathbf R,\mathbf P)={}&-\Re\Big\{\sum_{n,m}A_{n,m}\,C(n,m)\,\beta_0(n,m)\,e^{i\mathbf G\cdot\mathbf P}\Big\}+V(\mathbf R)\\
&+\frac{1}{2\omega^2}\Big\langle\boldsymbol\xi_1^{T}(\tau;\mathbf P)\,\mathbf H(\mathbf R)\,\boldsymbol\xi_1(\tau;\mathbf P)\Big\rangle_\tau,
\end{split}
\end{equation}
with $\mathbf G=na\,\mathbf i+mb\,\mathbf j$, $\mathbf H(\mathbf R)=\nabla\nabla V(\mathbf R)$, $C(n,m)=\exp\!\big(i\sum_{k\neq0}\alpha_k\big)$, $\alpha_k=\mathbf G\cdot\mathbf f_k/(ik\omega)$, and $\beta_\ell(n,m)$ are the Fourier coefficients of $F(\tau)=\exp\!\big(-i\sum_{k\neq0}\alpha_ke^{ik\tau}\big)$, which hold under the hypothesis $\mathbf f_0=\mathbf 0$. The key identity is equation \eqref{eq:prome3}: $\big\langle e^{-i\mathbf G\cdot\mathbf g(t)}\big\rangle=C\beta_0$, with
$\mathbf g(t)=\int_0^t\mathbf f(\omega t')dt'$.

In this case, let us consider the components of the oscillating field of the form:
\begin{equation}
f_x(t)=
\begin{cases}
\dfrac{f_1}{\sqrt{49/16-6t/\tau}}, & 0<t<\tau/2,\\[6pt]
-\dfrac{f_1}{\sqrt{49/16-6(t/\tau-1/2)}}, & \tau/2<t<\tau,
\end{cases}\label{eq:fxdef}
\end{equation}
\begin{equation}
f_y(t)=
\begin{cases}
\dfrac{f_2}{\sqrt{49/16-6t/\tau}}, & 0<t<\tau/2,\\[6pt]
-\dfrac{f_2}{\sqrt{49/16-6(t/\tau-1/2)}}, & \tau/2<t<\tau.
\end{cases}
\label{eq:fydef}
\end{equation}
Comparing both expressions \eqref{eq:fxdef} and \eqref{eq:fydef}, one verifies that $f_x(t)$ and $f_y(t)$ have \textbf{exactly the same time dependence}, differing only in the constant real multiplicative factor $f_1$ versus $f_2$. The arguments of the square roots are identical to each other and use $t/\tau$ (as a dimensionless variable) in both segments. This allows us to \textbf{factorize} the time dependence:
\begin{equation}
f_x(t)=f_1\,s(t),\quad f_y(t)=f_2\,s(t),\quad s(t)\equiv
\begin{cases}
\dfrac1{\sqrt{49/16-6t/\tau}}, & 0<t<\tau/2,\\[6pt]
-\dfrac1{\sqrt{49/16-6(t/\tau-1/2)}}, & \tau/2<t<\tau.
\end{cases}
\label{eq:sdef}
\end{equation}
This factorization is the central idea for the two-dimensional treatment of this case, where it can be noted that the vector field does not rotate in the $xy$ plane, but rather \textbf{oscillates along a fixed direction} $f_1\mathbf i+f_2\mathbf j$, with a common scalar envelope $s(t)$.

\subsubsection{Properties of $s(t)$: periodicity and symmetry}

\textbf{Domain.} For the interval $0<t<\tau/2$, $t/\tau\in(0,1/2)\Rightarrow6t/\tau\in(0,3)$, so the argument of the square root in the denominator has domain $\in(1/16,49/16)$, which is always positive, thereby avoiding any inconsistency; analogously, a similar analysis in the second segment shows that the root is well defined
at every interior point of each interval.

\textbf{Notable values.} There are jump discontinuities at the points $t=\tau/2$ and at $t=0\equiv\tau$, of equal magnitude $32/7$ and opposite sign, thus:
\begin{equation}
s(0^+)=\frac4 7,\qquad s\!\left(\frac\tau2^-\right)=4,\qquad s\!\left(\frac\tau2^+\right)=-\frac47,\qquad s(\tau^-)=-4.
\end{equation}
\begin{figure}[h]
\centering
\hspace{0em}\includegraphics[scale=0.60,angle=0]{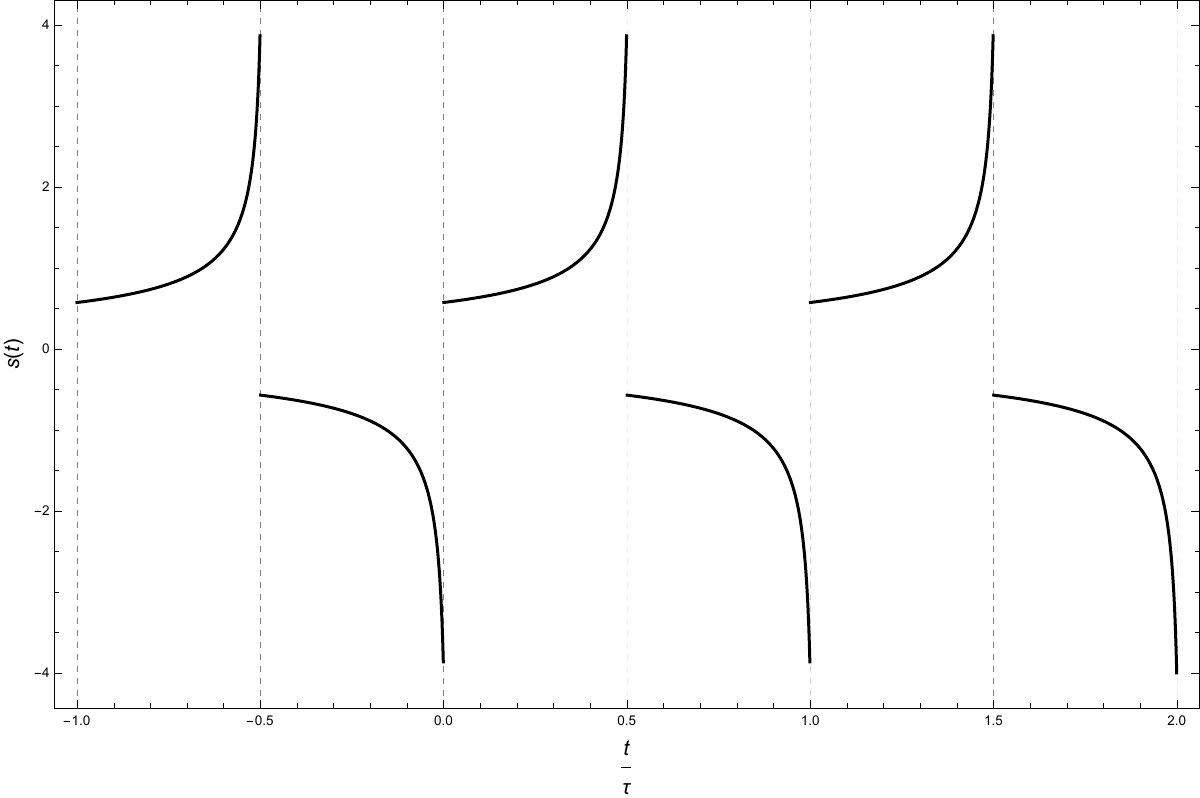}
\caption{\label{fig1}Plot of $s(t)$}
\end{figure}

\textbf{Half-wave symmetry.} For $t\in(0,\tau/2)$, evaluating the second segment at $t+\tau/2$:
\begin{equation}
s\!\left(t+\frac\tau2\right)=-\frac1{\sqrt{49/16-6\big((t+\tau/2)/\tau-1/2\big)}}=-\frac1{\sqrt{49/16-6t/\tau}}=-s(t).
\label{eq:santisim}
\end{equation}
Hence $s(t+\tau/2)=-s(t)$, so it has half-wave symmetry. Since $f_x(t)=f_1s(t)$ and $f_y(t)=f_2s(t)$, \textbf{both components inherit the same symmetry}, that is:
\begin{equation}
f_x\!\left(t+\frac\tau2\right)=-f_x(t),\qquad f_y\!\left(t+\frac\tau2\right)=-f_y(t).
\label{eq:zxzyantisim}
\end{equation}

\textbf{Consequence for the Fourier coefficients.} Writing the Fourier sum for $s(t)=\sum_ks_ke^{ik\omega t}$. Splitting $\int_0^\tau$ into two halves,
\begin{align}
s_k&=\frac{1}{\tau}\int^{\tau}_0s(t)e^{-ik\omega t}dt=\frac{1}{\tau}\left(\int^{\tau/2}_0s(t)e^{-ik\omega t}dt+\int^{\tau}_{\tau/2}s(t)e^{-ik\omega t}dt\right)\nonumber\\
&=\frac{1-(-1)^k}\tau\int_0^{\tau/2}s(t)e^{-ik\omega t}dt\ \Longrightarrow\ s_k=0\ \text{for every even $k$, in particular }s_0=0,
\label{eq:skpar}
\end{align}
where in the integral on the right-hand side of \eqref{eq:skpar} the substitution $t\to t+\tau/2$ was carried out in the second part of the integral, and \eqref{eq:santisim} was then used, together with the substitution of the identity $e^{-ik\pi}=(-1)^k$. Since $f_{k,x}=f_1s_k$ and $f_{k,y}=f_2s_k$, which are the $x,y$ components of $\mathbf f_k$, it follows that $f_{0,x}=f_{0,y}=0$, i.e. $\mathbf f_0=\mathbf 0$, whereby the hypothesis used in Section~5 is \textbf{proven} for this type of field and for any value of $f_1$ and $f_2$.

\subsubsection{Zero static potential and field-dependent quantities in the effective Hamiltonian}

For the case $V(x,y)=0$, the quantities that explicitly depend on the oscillating field in the final effective Hamiltonian \eqref{eq:Heffinal} are $C(n,m)$ and $\beta_0(n,m)$, whose product is computed directly via the equation:
\begin{equation}
C(n,m)\beta_0(n,m)=\big\langle e^{-i\mathbf G\cdot\mathbf g(t)}\big\rangle_t=\frac1\tau\int_0^\tau e^{-i(nag_x(t)+mbg_y(t))}dt,
\label{eq:Cbeta0v3}
\end{equation}
with $g_x(t)=\int_0^tf_x(t')dt'$, $g_y(t)=\int_0^tf_y(t')dt'$, the $x,y$ components of $\mathbf g(t)$.

\textbf{Computation of $g_x(t)$ and $g_y(t)$:} Using equation \eqref{eq:sdef} and factorizing,
\begin{equation}
\begin{split}
g_x(t)&=\int_0^tf_1s(t')dt'=f_1\,S(t),\qquad g_y(t)=\int_0^tf_2s(t')dt'=f_2\,S(t),\\
S(t)&\equiv\int_0^ts(t')dt'.
\end{split}
\label{eq:gxgyfact}
\end{equation}
\textbf{Computation of $S(t)$, for the segment $0\le t\le\tau/2$:} let the change of variables be $u=49/16-6t'/\tau$, so $du=-6/\tau\,dt'$; evaluating at $t'=0\Rightarrow u=49/16$ and at $t'=t\Rightarrow u=49/16-6t/\tau$:
\begin{equation}
S(t)=\int_0^t\frac{dt'}{\sqrt{49/16-6t'/\tau}}=-\frac\tau6\Big[2\sqrt u\Big]_{49/16}^{49/16-6t/\tau}
=\frac\tau3\left[\frac74-\sqrt{\frac{49}{16}-\frac{6t}\tau}\right].
\label{eq:S1}
\end{equation}
Evaluating at the endpoints of the interval analyzed: at $t=0$, $S(0)=0$; and at $t=\tau/2$, note that $\sqrt{49/16-3}=1/4\Rightarrow S(\tau/2)=\dfrac\tau3\big(\dfrac74-\dfrac14\big)=\dfrac\tau2$.

\textbf{For the segment $\tau/2\le t\le\tau$:} $S(t)=S(\tau/2)+\int_{\tau/2}^ts(t')dt'$; with the change of variables: $v=49/16-6(t'/\tau-1/2)$, $dv=-6/\tau\,dt'$, and a calculation similar to the previous case:
\begin{equation}
S(t)=\frac\tau2-\frac\tau3\left[\frac74-\sqrt{\frac{49}{16}-6\Big(\frac t\tau-\frac12\Big)}\right].
\label{eq:S2}
\end{equation}
Then, at $t=\tau$: $\sqrt{49/16-3}=1/4$, the bracket $=7/4-1/4=3/2$, $S(\tau)=\dfrac\tau2-\dfrac\tau3\cdot\dfrac32=\dfrac\tau2-\dfrac\tau2=0=S(0)$, so \textbf{$S(t)$ is periodic}.

\subsubsection{Linear combination $nag_x(t)+mbg_y(t)$}

From equation \eqref{eq:gxgyfact},
\begin{equation}
n a\,g_x(t)+mb\,g_y(t)=(naf_1+mbf_2)\,S(t).
\label{eq:combolineal}
\end{equation}
This is the explicit algebraic appearance of the combination $naf_1+mbf_2$, arising directly from the dot product $\mathbf G\cdot\mathbf g(t)$ upon factorizing $g_x=f_1S$ and $g_y=f_2S$. If we define, purely for notational convenience,
\begin{equation}
\Lambda_{n,m}\equiv(naf_1+mbf_2)\,\tau=\frac{2\pi(naf_1+mbf_2)}\omega.
\label{eq:Lambdanm}
\end{equation}
where we explicitly introduce the phase variable $\theta=\omega t$ and $\Gamma(\theta)\equiv S(\theta)/\tau$, so that $S(t)=\tau\Gamma(\theta)$, then, with the previous definitions:
\begin{equation}
(naf_1+mbf_2)S(t)=(naf_1+mbf_2)\tau\frac{S(t)}{\tau}=\Lambda_{n,m}\,\Gamma(\theta).
\label{eq:fasefinal}
\end{equation}
where $\Gamma(\theta)$ is defined by:
\begin{equation}
\Gamma(\theta)=
\begin{cases}
\dfrac13\left[\dfrac74-\sqrt{\dfrac{49}{16}-\dfrac{3\theta}\pi}\right], & 0\le\theta\le\pi,\\[8pt]
\dfrac12-\dfrac13\left[\dfrac74-\sqrt{\dfrac{49}{16}-\dfrac{3\theta}\pi+3}\right], & \pi\le\theta\le2\pi.
\end{cases}
\label{eq:Gammathetaprev}
\end{equation}

\subsubsection{Computation of $C(n,m)\beta_0(n,m)$ and its reduction to elementary form}
Substituting~\eqref{eq:fasefinal} into~\eqref{eq:Cbeta0v3}, with $t/\tau=\omega t/(2\pi)=\theta/(2\pi)\rightarrow dt/d\tau=d\theta/(2\pi)$, we then get
\begin{equation}
C(n,m)\beta_0(n,m)=\frac1{2\pi}\int_0^{2\pi}e^{-i\Lambda_{n,m}\Gamma(\theta)}d\theta=\frac1{2\pi}(I_1+I_2)\equiv\Phi(\Lambda_{n,m}),
\label{eq:PhiLambdanm}
\end{equation}
where
\begin{equation}
I_1=\int_0^\pi e^{-i\Lambda_{n,m}\Gamma(\theta)}d\theta,\quad I_2=\int_\pi^{2\pi}e^{-i\Lambda_{n,m}\Gamma(\theta)}d\theta.
\label{eq:PhiDefnew}
\end{equation}

\textbf{Computation of $I_1$.} Let the change of variables be $u=\sqrt{49/16-3\theta/\pi}$, from which we solve $\theta=\dfrac\pi3\Big(\dfrac{49}{16}-u^2\Big)$, whose differential is $d\theta=-\dfrac{2\pi u}3\,du$. The limits change as: $\theta=0\Rightarrow u=7/4$; $\theta=\pi\Rightarrow u=1/4$. Thus:
\begin{align}
I_1&=\int_{u=7/4}^{1/4}e^{-i\Lambda_{n,m}\frac13(7/4-u)}\left(-\frac{2\pi u}3\right)du
=\frac{2\pi}3\int_{1/4}^{7/4}u\,e^{-i\Lambda_{n,m}(7/4-u)/3}\,du\nonumber\\
&=\frac{2\pi}3\,e^{-7i\Lambda_{n,m}/12}\int_{1/4}^{7/4}u\,e^{ic u}\,du,
\label{eq:I1intnew}
\end{align}
where $c\equiv\Lambda_{n,m}/3$ is defined. Integrating the integral by parts: $\displaystyle\int u\,e^{icu}\,du=e^{icu}\Big(\frac1{c^2}-\frac{iu}c\Big)$. Evaluating at the lower limit $u=1/4$ and upper limit $u=7/4$, with $7c/4=7\Lambda_{n,m}/12$, $c/4=\Lambda_{n,m}/12$:
\begin{equation}
\int_{1/4}^{7/4}u\,e^{icu}\,du=e^{7i\Lambda_{n,m}/12}\Big(\frac1{c^2}-\frac{7i}{4c}\Big)-e^{i\Lambda_{n,m}/12}\Big(\frac1{c^2}-\frac i{4c}\Big).
\end{equation}
Substituting into \eqref{eq:I1intnew} and simplifying:
\begin{equation}
I_1=\frac{2\pi}3\left[\Big(\frac1{c^2}-\frac{7i}{4c}\Big)-e^{-i\Lambda_{n,m}/2}\Big(\frac1{c^2}-\frac i{4c}\Big)\right],
\label{eq:I1resnew}
\end{equation}

\textbf{Computation of $I_2$.} To compute the integral more efficiently, we perform the following change of variables: $\theta=\theta'+\pi$, where the new variable ranges over $\theta'\in[0,\pi]$; from equation \eqref{eq:Gammathetaprev} for the second part of the integral, one verifies that $\Gamma_2(\theta'+\pi)=\tfrac12-\Gamma_1(\theta')$, where the equivalence for the radicand $49/16-3\theta/\pi+3=49/16-3\theta'/\pi$ was used. Therefore
\begin{equation}
I_2=\int_\pi^{2\pi}e^{-i\Lambda_{n,m}\Gamma_2(\theta)}d\theta=\int_0^\pi e^{-i\Lambda_{n,m}\left[\frac12-\Gamma_1(\theta')\right]}d\theta'=e^{-i\Lambda_{n,m}/2}\int_0^\pi e^{i\Lambda_{n,m}\Gamma_1(\theta')}d\theta'.
\label{eq:I2setup}
\end{equation}
The remaining integral equals $I_1$ with the substitution $\Lambda_{n,m}\to-\Lambda_{n,m}$, which also induces the sign change $c\to-c$ according to its definition in each exponent. Repeating exactly the previous calculation, and with $\int u\,e^{-icu}du=e^{-icu}\big(\tfrac1{c^2}+\tfrac{iu}c\big)$, we have:
\begin{equation}
\int_0^\pi e^{i\Lambda_n\Gamma_1(\theta')}d\theta'=\frac{2\pi}3\left[\Big(\frac1{c^2}+\frac{7i}{4c}\Big)-e^{i\Lambda_{n,m}/2}\Big(\frac1{c^2}+\frac i{4c}\Big)\right].\label{eq:inti1}
\end{equation}
Substituting~\eqref{eq:inti1} into~\eqref{eq:I2setup} and simplifying:
\begin{equation}
I_2=\frac{2\pi}3\left[e^{-i\Lambda_{n,m}/2}\Big(\frac1{c^2}+\frac{7i}{4c}\Big)-\Big(\frac1{c^2}+\frac i{4c}\Big)\right].
\label{eq:I2resnew}
\end{equation}

Adding and simplifying~\eqref{eq:I1resnew} and~\eqref{eq:I2resnew}, one can see that the $1/c^2$ terms cancel exactly, so that
\begin{equation}
I_1+I_2=\frac{2\pi}3\left[-\frac{2i}c+e^{-i\Lambda_{n,m}/2}\frac{2i}c\right]=\frac{4\pi i}{3c}\big(e^{-i\Lambda_{n,m}/2}-1\big)=\frac{4\pi i}{\Lambda_{n,m}}\big(e^{-i\Lambda_{n,m}/2}-1\big),
\end{equation}
using $c=\Lambda_{n,m}/3$. Thus, from~\eqref{eq:PhiLambdanm}:
\begin{equation}
\Phi(\Lambda_{n,m})=\frac{2i}{\Lambda_{n,m}}\big(e^{-i\Lambda_{n,m}/2}-1\big).\label{eq:Philam}
\end{equation}
Factoring
\[
e^{-i\Lambda_{n,m}/2}-1=e^{-i\Lambda_{n,m}/4}\big(e^{-i\Lambda_{n,m}/4}-e^{i\Lambda_{n,m}/4}\big)=-2i\,e^{-i\Lambda_{n,m}/4}\sin(\Lambda_{n,m}/4),
\]
Eq.~\eqref{eq:Philam} finally takes the compact form
\begin{equation}
\Phi(\Lambda_{n,m})\equiv C(n,m)\beta_0(n,m)=\frac{4\sin(\Lambda_{n,m}/4)}{\Lambda_{n,m}}\,e^{-i\Lambda_{n,m}/4}.\label{eq:Phifinalnew}
\end{equation}
Note that $\Phi(0)=1$, whose modulus is $|\Phi(\Lambda)|=|\mathrm{sinc}(\Lambda/4)|\le1$; moreover $\Phi(-\Lambda)=\overline{\Phi(\Lambda)}$, since recall that $\Phi(\Lambda)=\langle e^{-i\Lambda\Gamma(\theta)}\rangle$ with $\Gamma$ real.

\subsubsection{Separation by hopping element: $A_n$, $B_m$, $C_{n,m}$}

According to the reflection symmetry discussed in Section~1, $A_{\pm n,0}=A_n$ (pure hopping in the $x$ direction), $A_{0,\pm m}=B_m$ (pure hopping in the $y$ direction), and $A_{\pm n,\pm m}=C_{n,m}$ for $n,m\neq0$ (the diagonal-direction hoppings, with the same value for the four sign combinations). Substituting~\eqref{eq:Phifinalnew} into~\eqref{eq:Heffinal} with $V(x,y)=0$ and
separating the double sum into the $x$, $y$, and diagonal directions, we obtain:
\begin{align}
H_{\mathrm{ef}}&=-\Re\Big\{\sum_{n\neq0}A_n\,\Phi(\Lambda_{n,0})\,e^{inaP_x}\Big\}-\Re\Big\{\sum_{m\neq0}B_m\,\Phi(\Lambda_{0,m})\,e^{imbP_y}\Big\}\nonumber\\
&-\Re\Big\{\sum_{n,m\neq0}C_{n,m}\,\Phi\big(\Lambda_{n,m}\big)\,e^{i(naP_x+mbP_y)}\Big\}. \label{eq:Hsep}
\end{align}

\textbf{Pure hopping in the $x$ direction}: Recalling that $\Lambda_{-n,0}=-\Lambda_{n,0}$, $\Phi(-\Lambda)=\overline{\Phi(\Lambda)}$, and $A_{-n}=A_n$ real, the terms in $n$ and $-n$ are complex conjugates and combine into a cosine:
\begin{equation}
-\Re\Big\{\sum_{n\neq0}A_n\Phi(naf_1\tau)e^{inaP_x}\Big\}=-2\sum_{n>0}A_n\,\frac{4\sin(naf_1\tau/4)}{naf_1\tau}\cos\!\left(naP_x-\frac{naf_1\tau}4\right).
\label{eq:Hx}
\end{equation}

\textbf{Pure hopping in the $y$ direction}: Identical to the previous case, with the substitution $m\leftrightarrow n$, $b\leftrightarrow a$, $f_2\leftrightarrow f_1$:
\begin{equation}
-\Re\Big\{\sum_{m\neq0}B_m\Phi(mbf_2\tau)e^{imbP_y}\Big\}=-2\sum_{m>0}B_m\,\frac{4\sin(mbf_2\tau/4)}{mbf_2\tau}\cos\!\left(mbP_y-\frac{mbf_2\tau}4\right).
\label{eq:Hy}
\end{equation}

\textbf{Pure hopping in the diagonal direction}: The sum $\sum_{n,m\neq0}$ runs over the four index quadrants. Fixing $n,m>0$, the four sign combinations are grouped and, using the property $C_{n,m}=C_{-n,-m}=C_{n,-m}=C_{-n,m}$, we have:
\begin{align}
\Lambda_{n,m}&=(naf_1+mbf_2)\tau\equiv\Lambda^+_{n,m},\quad\Rightarrow &\Lambda_{-n,-m}&=-\Lambda^+_{n,m},\\
\Lambda_{n,-m}&=(naf_1-mbf_2)\tau\equiv\Lambda^-_{n,m},\quad\Rightarrow&\Lambda_{-n,m}&=-\Lambda^-_{n,m}.
\end{align}
Here the linear combination between the lattice constants in the $x$ and $y$ directions and the scalar coefficients $f_1$ and $f_2$ appears explicitly, and without being introduced arbitrarily, in the form $naf_1\pm mbf_2$. In general, for arbitrary $n,m$,
\begin{equation}
\Lambda^+_{n,m}=(naf_1+mbf_2)\tau,\qquad \Lambda^-_{n,m}=(naf_1-mbf_2)\tau.
\end{equation}
Using the already proven property $\Phi(-\Lambda)=\overline{\Phi(\Lambda)}$ and, for the diagonal hopping elements, $C_{n,m}\in\mathbb R$, the term corresponding to the diagonal hoppings reduces to:
\begin{align}
&C_{n,m}\Phi(\Lambda^+_{n,m})e^{i(naP_x+mbP_y)}+C_{n,m}\Phi(-\Lambda^+_{n,m})e^{-i(naP_x+mbP_y)}\nonumber\\
&\qquad=2C_{n,m}\Re\Big\{\Phi(\Lambda^+_{n,m})e^{i(naP_x+mbP_y)}\Big\},\\
&C_{n,m}\Phi(\Lambda^-_{n,m})e^{i(naP_x-mbP_y)}+C_{n,m}\Phi(-\Lambda^-_{n,m})e^{-i(naP_x-mbP_y)}\nonumber\\
&\qquad=2C_{n,m}\Re\Big\{\Phi(\Lambda^-_{n,m})e^{i(naP_x-mbP_y)}\Big\}.
\end{align}
Therefore, using Eq.~\eqref{eq:Phifinalnew} in $\Re\{\Phi(\Lambda)e^{i\varphi}\}=\dfrac{4\sin(\Lambda/4)}{\Lambda}\cos(\varphi-\Lambda/4)$:
\begin{align}
&-\Re\Big\{\sum_{n,m\neq0}C_{n,m}\Phi(\Lambda_{n,m})e^{i(naP_x+mbP_y)}\Big\}\nonumber\\
&\quad=-2\sum_{n,m>0}C_{n,m}\,\frac{4\sin(\Lambda^+_{n,m}/4)}{\Lambda^+_{n,m}}\cos\!\left(naP_x+mbP_y-\frac{\Lambda^+_{n,m}}4\right)\nonumber\\
&\qquad-2\sum_{n,m>0}C_{n,m}\,\frac{4\sin(\Lambda^-_{n,m}/4)}{\Lambda^-_{n,m}}\cos\!\left(naP_x-mbP_y-\frac{\Lambda^-_{n,m}}4\right).
\label{eq:Hdiag}
\end{align}

\subsubsection{Final effective Hamiltonian}

Finally, adding Eqs.~\eqref{eq:Hx}, \eqref{eq:Hy}, and \eqref{eq:Hdiag} we obtain:
\begin{equation}
\begin{aligned}
H_{\mathrm{ef}}(P_x,P_y)=&-2\sum_{n>0}A_n\,\frac{4\sin(\Lambda^x_n/4)}{\Lambda^x_n}\cos\!\left(naP_x-\frac{\Lambda^x_n}4\right)\\
&-2\sum_{m>0}B_m\,\frac{4\sin(\Lambda^y_m/4)}{\Lambda^y_m}\cos\!\left(mbP_y-\frac{\Lambda^y_m}4\right)\\
&-2\sum_{n,m>0}C_{n,m}\left[\frac{4\sin(\Lambda^+_{n,m}/4)}{\Lambda^+_{n,m}}\cos\!\left(naP_x+mbP_y-\frac{\Lambda^+_{n,m}}4\right)\right.\\
&\left.\hspace{2.7cm}+\frac{4\sin(\Lambda^-_{n,m}/4)}{\Lambda^-_{n,m}}\cos\!\left(naP_x-mbP_y-\frac{\Lambda^-_{n,m}}4\right)\right]
\end{aligned}, \label{eq:Hfinal2D}
\end{equation}
with
\begin{equation}
\Lambda^x_n=naf_1\tau,\qquad \Lambda^y_m=mbf_2\tau,\qquad \Lambda^\pm_{n,m}=(naf_1\pm mbf_2)\tau=\frac{2\pi(naf_1\pm mbf_2)}\omega.
\label{eq:Lambdasdef}
\end{equation}

\subsubsection{Some preliminary remarks}

\textbf{Periodicity ($\omega\tau=2\pi$).} $\Lambda^x_n,\Lambda^y_m,\Lambda^\pm_{n,m}$ do not depend on
$t$, since they are constants after time averaging over $T=2\pi/\omega=\tau$, as required by the use of $\langle\cdot\rangle_\tau$ in Eq.~\eqref{eq:Heffinal}. Moreover $H_{\mathrm{ef}}^{(V=0)}(P_x+2\pi/a,P_y)=H_{\mathrm{ef}}^{(V=0)}(P_x,P_y+2\pi/b)=H_{\mathrm{ef}}^{(V=0)}(P_x,P_y)$, so the periodicity of the first Brillouin zone holds in both directions.

\textbf{Symmetries.} The half-wave symmetry~\eqref{eq:santisim}, common to $f_x$ and $f_y$ since they
share $s(t)$, produced $\mathbf f_0=\mathbf0$ with the help of Eq.~\eqref{eq:skpar}, which is a necessary condition for applying Eq.~\eqref{eq:prome3} as such. The reflection symmetry $A_{\pm n,\pm m}=C_{n,m}$ for the diagonal hopping elements, together with the property $\Phi(-\Lambda)=\overline{\Phi(\Lambda)}$, a consequence of $\Gamma(\theta)$ being real, guaranteed that each block of four index terms reduced to two real cosines, which was quite helpful for the final simplification.

\textbf{Cross-checks.} If $f_1\to0,f_2\to0$, equivalent to taking $\omega\to\infty$ for fixed $f_1$ and $f_2$ (since
$\Lambda\propto1/\omega$), all $\Lambda\to0$, $\Phi\to1$, and Eq.~\eqref{eq:Hfinal2D} reduces to
$-2\sum_{n>0}A_n\cos(naP_x)-2\sum_{m>0}B_m\cos(mbP_y)-2\sum_{n,m>0}C_{n,m}\big[\cos(naP_x+mbP_y)+\cos(naP_x-mbP_y)\big]$,
which is exactly $-\Re\{\sum_{n,m}A_{n,m}e^{i\mathbf G\cdot\mathbf P}\}$, the unperturbed two-dimensional tight-binding band, consistent with setting $\mathbf f\equiv\mathbf 0$ directly in Eq.~\eqref{eq:Heffinal} ($C=\beta_0=1$).

The oscillating field $\mathbf f(\omega t)=s(t)(f_1\mathbf i+f_2\mathbf j)$ considered here is a \emph{linearly polarized} oscillating field, non-rotating in the $xy$ plane, with fixed direction $(f_1,f_2)$ and common temporal envelope $s(t)$. Its effect on the two-dimensional tight-binding band in the absence of the static potential is to independently renormalize the three types of hopping: the pure hoppings in $x$ see only the effect of $f_1$ through $\Lambda^x_n$, the pure hoppings in $y$ see only the effect of $f_2$ through $\Lambda^y_m$, and the diagonal hoppings see the effect of \textbf{both} components, explicitly displaying the combinations $f_1+f_2$ — interpretable as a ``co-diagonal'' channel, through the term $\Lambda^+$ — and $f_1-f_2$ — interpretable as an ``anti-diagonal'' channel, through the term $\Lambda^-$.

\subsubsection{Anisotropic band collapse and absence of exact dynamic localization}

From Eqs.~\eqref{eq:Hfinal2D} and \eqref{eq:Phifinalnew}, the renormalization factor of the hopping elements is
\[
A_{n,m}C(n,m)\beta_0(n,m)=A_{n,m}\Phi(\Lambda_{n,m}),
\]
where, in general:
\begin{equation}
\Phi(\Lambda)=\frac{4\sin(\Lambda/4)}\Lambda\,e^{-i\Lambda/4}.
\label{eq:PhiFin}
\end{equation}

From~\eqref{eq:PhiFin}, the exact vanishing condition of $\Phi(\Lambda)$ leads to:
\begin{equation}
\Phi(\Lambda)=0\iff\sin\!\left(\frac\Lambda4\right)=0\ \text{and}\ \Lambda\neq0\iff\Lambda=4\pi j,\quad j\in\mathbb Z\setminus\{0\}.
\label{eq:condexactaFin}
\end{equation}
Here the case $j=0$, equivalent to $\Lambda=0$, is explicitly excluded, since in that case $\Phi(0)=\lim_{\Lambda\to0}4\sin(\Lambda/4)/\Lambda=1\neq0$, where the well-known limit $\lim_{x\to0}\sin x/x\to1$ was used, which does not lead to a vanishing in the strict sense.

Next, the condition for band suppression in the $x$ and $y$ directions is imposed; from Eq.~\eqref{eq:Hfinal2D} one can see that this is achieved if:
\begin{equation}
\frac{naf_1\tau}4=\pi k,\qquad \frac{mbf_2\tau}4=\pi l,\qquad \forall k,l\in\mathbb Z\setminus\{0\},\quad l\neq k.
\label{eq:condpropuesta}
\end{equation}
For the first equation of~\eqref{eq:condpropuesta} to make sense as a \emph{single} condition, valid simultaneously for every value of $n$, the numerical value of $k$ cannot be a free variable re-evaluated for each $n$; it must instead be a fixed constant, corresponding to the case $n=1$, so that for a general $n$ one has:
\begin{equation}
\frac{naf_1\tau}4=n\left(\frac{af_1\tau}4\right)=n\pi k=\pi(nk),
\label{eq:nkml}
\end{equation}
and, performing a similar analysis for the other direction,
\begin{equation}
\frac{mbf_2\tau}4=\pi(ml).
\label{eq:nkml1}
\end{equation}
Substituting~\eqref{eq:nkml} and~\eqref{eq:nkml1} into the definition of $\Lambda^\pm_{n,m}/4$ given in~\eqref{eq:Lambdasdef}, we obtain:
\begin{equation}
\frac{\Lambda^\pm_{n,m}}4=\frac{naf_1\tau}4\pm\frac{mbf_2\tau}4=\pi(nk)\pm\pi(ml)=\pi(nk\pm ml)\equiv\pi p_\pm,
\label{eq:LambdapmVerif}
\end{equation}
where $p_\pm\equiv nk\pm ml\in\mathbb Z$ is defined, an integer as a combination of integers.

On the other hand, the vanishing condition~\eqref{eq:condexactaFin} requires $\Lambda\neq0$, i.e., $p_\pm\neq0$, not merely $p_\pm\in\mathbb Z$. For the unidirectional sectors: in $x$, $p^x=nk$ with $n\geq1$, $k\neq0$ fixed $\Rightarrow p^x\neq0$ always and without exception; analogously for the $y$ direction, where $p^y=ml\neq0$. For the diagonal direction, however, $p_-=nk-ml$ or $p_+=nk+ml$ \emph{can} be made to vanish for particular pairs $(n,m)$ with $n,m\geq1$: the Diophantine equation $nk=ml$ always has a positive integer solution, namely when the pair $(n,m)$ takes the value $(l,k)$ — taking $k,l>0$ without loss of generality; for the general case the pair takes the value $(|l|,|k|)$ — since this yields $lk=kl$, satisfied trivially. More formally, if $d$ is the greatest common divisor of $(k,l)$, then $d$ divides both $k$ and $l$, so there exist integers $k'$ and $l'$ such that $k=dk'$, $l=dl'$, and moreover the greatest common divisor of $(k',l')$ necessarily equals $1$; the complete family of solutions is then
\begin{equation}
(n,m)=j\,(l',k'),\qquad j=1,2,3,\ldots
\label{eq:familiaresonanteFin}
\end{equation}
Note that for the pairs in~\eqref{eq:familiaresonanteFin}, the value $\Lambda^-_{n,m}=0$ exactly, and from Eq.~\eqref{eq:condexactaFin}, $\Phi(\Lambda^-_{n,m})=\Phi(0)=1\neq0$; thus the corresponding diagonal hopping element does \emph{not} vanish.

A possible restriction on the case described above might be to require $l\neq k$; however, this does not avoid this exception. This is verified explicitly by a counterexample: let $k=1,l=2$, with $l\neq k$ satisfying the additional hypothesis; the greatest common divisor of $(1,2)$ is $1$, so $(k',l')=(1,2)$, and the family of solutions according to~\eqref{eq:familiaresonanteFin} is $(n,m)=(2,1),(4,2),(6,3),\ldots$. At $(n,m)=(2,1)$:
\begin{equation}
\frac{\Lambda^-_{2,1}}4=\pi(2\cdot1-1\cdot2)=\pi(2-2)=0\ \Longrightarrow\ \Lambda^-_{2,1}=0,\quad \Phi(\Lambda^-_{2,1})=1\neq0.
\label{eq:contraejemploFin}
\end{equation}
However, the term corresponding to $\Lambda^+_{2,1}/4=\pi(2+2)=4\pi\Rightarrow\Phi(\Lambda^+_{2,1})=0$, consistent with $j=1\neq0$. Thus, the additional condition $l\neq k$ is mathematically \emph{irrelevant} for the total diagonal cancellation, since it guarantees it in no case, for any suitable choice of $k,l\in\mathbb Z\setminus\{0\}$.

In summary, the following can be stated:
\begin{itemize}
\item The renormalized hopping element in the $x$ direction, $A_n\,\Phi(\Lambda^x_n)=0$ for \emph{all} $n\geq1$.
\item The renormalized hopping element in the $y$ direction, $B_m\,\Phi(\Lambda^y_m)=0$ for \emph{all} $m\geq1$.
\item The renormalized hopping element in the $xy$ diagonal direction, $C_{n,m}\,\Phi(\Lambda^+_{n,m})=0$ for all $(n,m)$ with $n,m\geq1$, except when $nk=-ml$, possible only if $k,l$ have opposite signs.
\item The renormalized hopping element in the $xy$ diagonal direction, $C_{n,m}\,\Phi(\Lambda^-_{n,m})=0$ for all $(n,m)$ with $n,m\geq1$, except for the infinite family $(n,m)=j(l',k')$, $j=1,2,\ldots$, as described in Eq.~\eqref{eq:familiaresonanteFin}, where it equals $1$, not $0$.
\end{itemize}

For a two-dimensional oscillating field of the form $\mathbf f(\omega t)=s(t)(f_1\mathbf i+f_2\mathbf j)$ and in the absence of the static potential, $V(x,y)=0$, the quantization condition $af_1\tau/4=\pi k$, $bf_2\tau/4=\pi l$, with fixed $k,l\in\mathbb Z\setminus\{0\}$, the vanishing of the term in~\eqref{eq:PhiFin} is a necessary and sufficient condition to suppress the renormalized hopping element in the $x$ and $y$ crystallographic directions, and is moreover valid for \emph{every} neighbor order $n$ and $m$. However, this same condition does \textbf{not} cancel the renormalized diagonal hopping element for all $(n,m)$, since an infinite family of resonant neighbors subsists, as described by Eq.~\eqref{eq:familiaresonanteFin}, for which the renormalization factor $\Phi$ attains its maximum nonzero value. The additional condition $l\neq k$ imposed does not avoid this resonant family for any choice of nonzero integers $k,l$, as verified by the explicit counterexample $(n,m)=(2,1)$ with $k=1,l=2$.

Consequently, an \textbf{anisotropic collapse} of the band is achieved, since the band collapses completely in the pure $x$, $y$ directions and along most diagonal directions, but persists exactly along the residual commensurate direction. Thus, isotropic collapse in the strict sense is not attained, except in degenerate cases outside the two-dimensional regime — namely a unidirectional field, $f_1=0$ or $f_2=0$, or alternatively a lattice model without a diagonal hopping element, $C_{n,m}\equiv0$ — only in those cases would a total collapse of the band be achieved for every neighbor $(n,m)$.

It is further established that total band collapse for all neighbors $n,m$ — not only the anisotropic collapse obtained in this section — is mathematically \textbf{equivalent} to exact dynamic localization in the \emph{stroboscopic} sense of Dignam \cite{dignam2002}, i.e., the exact vanishing of the hopping elements between sites, with exact return of the wave packet at each period $T=\tau$; however, it must be stressed that this does \emph{not} guarantee the absence of wave-packet spreading at the intermediate instants $0<t<T$ within each period. Thus, since in the case described the collapse is anisotropic, \textbf{the exact dynamic localization condition is not guaranteed}, since the residual coupling along the resonant diagonal direction will prevent the exact stroboscopic return of the wave packet to its original site.

\subsection{Effective Hamiltonian for a field with mixed even--odd symmetry and an arbitrary static field}\label{sec:mixed}

\subsubsection{Oscillating field with mixed even--odd symmetry}
Let the oscillating field $\mathbf f(\omega t)$ have $x$ and $y$ components of the form:
\begin{equation}
\begin{split}
f_x(\omega t)&=2f_1\cos(\omega t)\quad\text{with even symmetry},\\
f_y(\omega t)&=2f_2\sin(\omega t)\quad\text{with odd symmetry}.
\end{split}
\label{eq:campoMixto}
\end{equation}
The Fourier coefficients of $\cos(\omega t)=\tfrac12(e^{i\omega t}+e^{-i\omega t})$ are $f_{x,\pm1}=f_1$, which is real, as in the even-symmetry case treated in a previous section. On the other hand, from the identity $\sin(\omega t)=\tfrac1{2i}(e^{i\omega t}-e^{-i\omega t})=-\tfrac{i}2(e^{i\omega t}-e^{-i\omega t})$, the Fourier coefficients are:
\begin{equation}
f_{y,1}=-if_2,\qquad f_{y,-1}=if_2,\label{eq:fycoef}
\end{equation}
which are purely imaginary, consistent with the general property already proven for odd-symmetry oscillating fields.

\subsubsection{Zeroth-order approximation in \texorpdfstring{$1/\omega$}{1/omega}}
From Eq.~\eqref{eq:Heffinal}, the curvature term is of order $\epsilon^2=1/\omega^2$ and is excluded for belonging to a different order of the expansion, not because it vanishes. Hence, what remains is:
\begin{equation}
H_{\mathrm{ef}}^{(0)}(\mathbf R,\mathbf P)=-\Re\Big\{\sum_{n,m}A_{n,m}C(n,m)\beta_0(n,m)e^{i\mathbf G\cdot\mathbf P}\Big\}+V(\mathbf R).
\label{eq:Hef0mixto}
\end{equation}

\subsubsection{Obtaining $\alpha_k(n,m)$}
Using the linearity already proven in a previous section, $\alpha_k(n,m)=n\alpha_k(1,0)+m\alpha_k(0,1)$, where $\alpha_k(1,0)=af_{x,k}/(ik\omega)$ and $\alpha_k(0,1)=bf_{y,k}/(ik\omega)$:
\begin{align}
\alpha_1(1,0)&=\frac{af_1}{i\omega}=-i\lambda_x,\qquad \alpha_{-1}(1,0)=\frac{af_1}{-i\omega}=i\lambda_x,\label{eq:alfax}\\
\alpha_1(0,1)&=\frac{b(-if_2)}{i\omega}=-\frac{bf_2}\omega=-\lambda_y,\qquad \alpha_{-1}(0,1)&=\frac{b(if_2)}{-i\omega}=-\frac{bf_2}\omega=-\lambda_y,\label{eq:alfay}
\end{align}
where $\lambda_x\equiv af_1/\omega$ and $\lambda_y\equiv bf_2/\omega$ are defined. Note that $\alpha_1(0,1)=\alpha_{-1}(0,1)=-\lambda_y$ is real and even in $k$, whereas the behavior of $\alpha_k(1,0)$ is purely imaginary and odd in $k$, consistent with the general property already proven in previous sections: even field $\Rightarrow\alpha_k$ imaginary and antisymmetric in $k$; odd field $\Rightarrow\alpha_k$ real and symmetric in $k$.

Therefore, for the general $(n,m)$ coefficients we have:
\begin{align}
\alpha_1(n,m)&=n(-i\lambda_x)+m(-\lambda_y)=-in\lambda_x-m\lambda_y,\label{eq:alfa1nm}\\
\alpha_{-1}(n,m)&=n(i\lambda_x)+m(-\lambda_y)=in\lambda_x-m\lambda_y.\label{eq:alfam1nm}
\end{align}

\subsubsection{The phase factor $C(n,m)$}
Section~5 defines the factor $C(n,m)$, which applied to the case of an oscillating field with mixed symmetry yields:
\begin{equation}
\begin{split}
C(n,m)&=\exp\big(i[\alpha_1(n,m)+\alpha_{-1}(n,m)]\big)\\
&=\exp\big(i[(-in\lambda_x-m\lambda_y)+(in\lambda_x-m\lambda_y)]\big)=\exp(-2im\lambda_y),
\end{split}
\label{eq:Cmixto}
\end{equation}
where, unlike the case of an even-symmetry field $\mathbf f(\omega t)$, for which $C=1$ always, here we obtain $C(n,m)=e^{-2im\lambda_y}$, a phase depending only on $m$, the index associated with the odd component of the field $\mathbf f(\omega t)$; note that the terms with $m=0$, corresponding to pure hoppings in $x$, have $C=1$.

\subsubsection{Derivation of $\beta_0(n,m)$}
First, the following expansion is simplified:
\begin{align}
\sum_{k\neq0}\alpha_k(n,m)e^{ik\tau}&=\alpha_1(n,m)e^{i\tau}+\alpha_{-1}(n,m)e^{-i\tau}\nonumber\\
&=(-in\lambda_x-m\lambda_y)e^{i\tau}+(in\lambda_x-m\lambda_y)e^{-i\tau}\nonumber\\
&=-in\lambda_x(e^{i\tau}-e^{-i\tau})-m\lambda_y(e^{i\tau}+e^{-i\tau})=2n\lambda_x\sin\tau-2m\lambda_y\cos\tau,
\label{eq:sumaMixta}
\end{align}
where Eqs.~\eqref{eq:alfa1nm}, \eqref{eq:alfam1nm}, and the identities $e^{i\tau}-e^{-i\tau}=2i\sin\tau$ and $e^{i\tau}+e^{-i\tau}=2\cos\tau$ were used. On the other hand, from the definition of $F(\tau;n,m)$ and using~\eqref{eq:sumaMixta}:
\begin{equation}
F(\tau;n,m)=\exp\big(-2in\lambda_x\sin\tau+2im\lambda_y\cos\tau\big)=e^{-2in\lambda_x\sin\tau}e^{2im\lambda_y\cos\tau}.
\label{eq:Fmixto}
\end{equation}
Applying the two forms of the Bessel generating identity:
\begin{equation}
e^{-2in\lambda_x\sin\tau}=\sum_pJ_p(-2n\lambda_x)e^{ip\tau}=\sum_p(-1)^pJ_p(2n\lambda_x)e^{ip\tau},
\label{eq:JAseno}
\end{equation}
\begin{equation}
e^{2im\lambda_y\cos\tau}=\sum_qi^qJ_q(2m\lambda_y)e^{iq\tau},
\label{eq:JAcoseno}
\end{equation}
where the Bessel function property $J_p(-x)=(-1)^pJ_p(x)$ was used in~\eqref{eq:JAseno}. Multiplying~\eqref{eq:JAseno} and~\eqref{eq:JAcoseno}:
\begin{equation}
F(\tau;n,m)=\sum_{p,q}(-1)^pJ_p(2n\lambda_x)i^qJ_q(2m\lambda_y)e^{i(p+q)\tau},
\label{eq:Fmixto1}
\end{equation}
defining the auxiliary variable $p+q=r$, we have:
\begin{equation}
F(\tau;n,m)=\sum_{r}\left[\sum_{p}(-1)^pJ_p(2n\lambda_x)i^{r-p}J_{r-p}(2m\lambda_y)\right]e^{ir\tau},
\label{eq:Fmixto2}
\end{equation}
and comparing~\eqref{eq:Fmixto2} with~\eqref{eq:Fp} we obtain:
\begin{equation}
\beta_{\ell}(n,m)=\sum_{p}(-1)^pJ_p(2n\lambda_x)i^{\ell-p}J_{\ell-p}(2m\lambda_y),
\label{eq:betafmixto}
\end{equation}
from which we directly obtain:
\begin{equation}
\beta_0(n,m)=\sum_{p}(-1)^pJ_p(2n\lambda_x)\,i^{-p}J_{-p}(2m\lambda_y).
\label{eq:beta0sumaPrevia}
\end{equation}
Applying the property $J_{-p}(2m\lambda_y)=(-1)^pJ_p(2m\lambda_y)$ to~\eqref{eq:beta0sumaPrevia} we have:
\begin{align}
\beta_0(n,m)&=\sum_p(-1)^p(-1)^p\,i^{-p}J_p(2n\lambda_x)J_p(2m\lambda_y)=\sum_p i^{-p}J_p(2n\lambda_x)J_p(2m\lambda_y)\nonumber\\
&=\sum_p(-i)^pJ_p(2n\lambda_x)J_p(2m\lambda_y).
\label{eq:beta0sumaBessel}
\end{align}

Using Graf's addition theorem for the zeroth-order case, from \cite{watson1944}, we have:
\begin{equation}
J_0\big(\sqrt{r^2+\rho^2-2r\rho\cos\varphi}\big)=\sum_{p=-\infty}^{\infty}J_p(r)J_p(\rho)\,e^{ip\varphi}.
\label{eq:Graf}
\end{equation}
For the particular case $\varphi=-\pi/2$, and using the identity
\[
e^{-ip\pi/2}=(e^{i\pi/2})^{-p}=(i)^{-p}=(-i)^p,
\]
then, performing the changes of variable $r=2n\lambda_x$, $\rho=2m\lambda_y$, and with $\cos(-\pi/2)=0$, we obtain:
\begin{equation}
\beta_0(n,m)=J_0\!\left(\sqrt{(2n\lambda_x)^2+(2m\lambda_y)^2}\right)=J_0\!\left(2\sqrt{n^2\lambda_x^2+m^2\lambda_y^2}\right).
\label{eq:beta0Mixto}
\end{equation}

\subsubsection{Case of a rectangular lattice with first- and second-neighbor interactions in the horizontal and vertical directions, and with a diagonal interaction}

Consider a rectangular lattice with lattice constants $a$ and $b$, with interactions up to second neighbors in the horizontal and vertical directions. The hopping elements corresponding to these interactions are characterized by $A_{1,0}=A_1$, $A_{2,0}=A_2$ in the $x$ direction, and by $A_{0,1}=B_1$, $A_{0,2}=B_2$ in the $y$ direction. Likewise, the diagonal interaction between first neighbors is considered, whose hopping element is given by $A_{1,1}=C_1$, as shown in Figure~\ref{fig:red-rectangular}.

In what follows, the contributions corresponding to each type of interaction are broken down systematically and explicitly. For each case, the expressions for $C(n,m)$, $\beta_0(n,m)$, and the term defined by $T_{n,m}\equiv A_{n,m}C(n,m)\beta_0(n,m)e^{i(naP_x+mbP_y)}$ are presented explicitly; this term represents the contribution associated with the effective hopping between sites separated by $na$ and $mb$ in the $x$ and $y$ directions. The breakdown of these contributions is summarized in Table~\ref{tab:interacciones-efectivas} below:

\begin{table}[H]
\captionsetup{justification=raggedright, singlelinecheck=false}
\caption{Effective interactions.}
\label{tab:interacciones-efectivas}
\centering
\begin{tabular}{|c|c|c|c|}
\hline
Case & $C(n,m)$ & $\beta_0(n,m)$ & $T_{n,m}$ \\
\hline
$n=1,\ m=0$ & 1 & $J_0(2\lambda_x)$ & $A_1\,J_0(2\lambda_x)\,e^{iaP_x}$ \\
\hline
$n=-1,\ m=0$ & 1&$J_0(2\lambda_x)$ & $A_1\,J_0(2\lambda_x)\,e^{-iaP_x}$\\
\hline
$n=2,\ m=0$ & 1&$J_0(4\lambda_x)$ & $A_2\,J_0(4\lambda_x)\,e^{2iaP_x}$\\
\hline
$n=-2,\ m=0$ & 1&$J_0(4\lambda_x)$ & $A_2\,J_0(4\lambda_x)\,e^{-2iaP_x}$\\
\hline
$n=0,\ m=1$ & $e^{-2i\lambda_y}$ & $J_0(2\lambda_y)$ & $B_1J_0(2\lambda_y)\,e^{i(bP_y-2\lambda_y)}$ \\
\hline
$n=0,\ m=-1$ & $e^{2i\lambda_y}$ & $J_0(2\lambda_y)$ & $B_1J_0(2\lambda_y)\,e^{-i(bP_y-2\lambda_y)}$ \\
\hline
$n=0,\ m=2$ & $e^{-4i\lambda_y}$ & $J_0(4\lambda_y)$ & $B_2J_0(4\lambda_y)\,e^{i(2bP_y-4\lambda_y)}$ \\
\hline
$n=0,\ m=-2$ & $e^{4i\lambda_y}$ & $J_0(4\lambda_y)$ & $B_2J_0(4\lambda_y)\,e^{-i(2bP_y-4\lambda_y)}$ \\
\hline
$n=1,\ m=1$ & $e^{-2i\lambda_y}$ & $J_0\left(2\sqrt{\lambda^2_x+\lambda^2_y}\right)$ & $C_{1}J_0\left(2\sqrt{\lambda^2_x+\lambda^2_y}\right)\,e^{i(aP_x+bP_y-2\lambda_y)}$ \\
\hline
$n=1,\ m=-1$ & $e^{2i\lambda_y}$ & $J_0\left(2\sqrt{\lambda^2_x+\lambda^2_y}\right)$ & $C_{1}J_0\left(2\sqrt{\lambda^2_x+\lambda^2_y}\right)\,e^{i(aP_x-bP_y+2\lambda_y)}$ \\
\hline
$n=-1,\ m=1$ & $e^{-2i\lambda_y}$ & $J_0\left(2\sqrt{\lambda^2_x+\lambda^2_y}\right)$ & $C_{1}J_0\left(2\sqrt{\lambda^2_x+\lambda^2_y}\right)\,e^{-i(aP_x-bP_y+2\lambda_y)}$ \\
\hline
$n=-1,\ m=-1$ & $e^{2i\lambda_y}$ & $J_0\left(2\sqrt{\lambda^2_x+\lambda^2_y}\right)$ & $C_{1}J_0\left(2\sqrt{\lambda^2_x+\lambda^2_y}\right)\,e^{-i(aP_x+bP_y-2\lambda_y)}$ \\
\hline
\end{tabular}
\end{table}

For the diagonal interaction, the symmetry property $A_{1,1}=A_{1,-1}=A_{-1,1}=A_{-1,-1}=C_1$ was used in Table~\ref{tab:interacciones-efectivas}. Likewise, the complex-conjugation relations $T_{-1,1}=\overline{T_{1,-1}}$ and $T_{-1,-1}=\overline{T_{1,1}}$ are verified.

Using the property $e^{i\theta}+e^{-i\theta}=2\cos\theta$ and the terms of the fourth column of Table~\ref{tab:interacciones-efectivas}:
\begin{align*}
T_{1,0}+T_{-1,0}&=2A_1J_0(2\lambda_x)\cos(aP_x),\nonumber\\
T_{2,0}+T_{-2,0}&=2A_2J_0(4\lambda_x)\cos(2aP_x),\nonumber\\
T_{0,1}+T_{0,-1}&=2B_1J_0(2\lambda_y)\cos(bP_y-2\lambda_y),\nonumber\\
T_{0,2}+T_{0,-2}&=2B_2J_0(4\lambda_y)\cos(2bP_y-4\lambda_y),\label{eq:par02}\\
T_{1,1}+T_{-1,-1}&=2C_{1}J_0\!\big(2\sqrt{\lambda_x^2+\lambda_y^2}\big)\cos(aP_x+bP_y-2\lambda_y),\nonumber\\
T_{1,-1}+T_{-1,1}&=2C_{1}J_0\!\big(2\sqrt{\lambda_x^2+\lambda_y^2}\big)\cos(aP_x-bP_y+2\lambda_y).\nonumber
\end{align*}

From the expressions $T_{1,1}+T_{-1,-1}$ and $T_{1,-1}+T_{-1,1}$, which are the contributions of the diagonal part, using the auxiliary definitions $\alpha\equiv aP_x$ and $\beta\equiv bP_y-2\lambda_y$, such that $aP_x+bP_y-2\lambda_y=\alpha+\beta$ and $aP_x-bP_y+2\lambda_y=\alpha-\beta$, the total contribution to the effective Hamiltonian of the diagonal part simplifies to:
\begin{equation}
H_{\mathrm{diag}}=2C_{1}J_0\!\big(2\sqrt{\lambda_x^2+\lambda_y^2}\big)\big[\cos(\alpha+\beta)+\cos(\alpha-\beta)\big]=4C_{1}J_0\!\big(2\sqrt{\lambda_x^2+\lambda_y^2}\big)\cos\alpha\cos\beta,
\label{eq:HdiagMixtoSimplificado}
\end{equation}
where the trigonometric identity $\cos(\alpha+\beta)+\cos(\alpha-\beta)=2\cos\alpha\cos\beta$ was used.

Finally, the expression for the zeroth-order effective Hamiltonian, taking into account the first- and second-neighbor interactions in the $x$, $y$ directions and the diagonal first-neighbor interaction, is:

\begin{align}
H_{\mathrm{ef}}^{(0)}(\mathbf R,\mathbf P)=&\ V(\mathbf R)-2A_1J_0(2\lambda_x)\cos(aP_x)-2A_2J_0(4\lambda_x)\cos(2aP_x)\nonumber\\
&-2B_1J_0(2\lambda_y)\cos(bP_y-2\lambda_y)-2B_2J_0(4\lambda_y)\cos(2bP_y-4\lambda_y)\nonumber\\
&-4C_{1}J_0\!\big(2\sqrt{\lambda_x^2+\lambda_y^2}\big)\cos(aP_x)\cos(bP_y-2\lambda_y),\label{eq:hef12diag}
\end{align}
where $\lambda_x=af_1/\omega$ and $\lambda_y=bf_2/\omega$. Note that in Eq.~\eqref{eq:hef12diag} every term with index $m\neq0$ exhibits the same substitution $bP_y\to bP_y-2m\lambda_y$, i.e., $P_y\to P_y-2\lambda_y/b$ for $m=1$ and $m=2$ as shown for the case described; that is, it is an effective shift of the crystal momentum in $y$ induced by the odd component of the oscillating field, entirely absent in the case of the even component of the oscillating field, where $\lambda_x$ only appears within the arguments of the Bessel function and not as a phase shift.

\subsubsection{A possible scenario for the duplication of the rectangular lattice}

From Eq.~\eqref{eq:hef12diag} for the zeroth-order effective Hamiltonian obtained, we investigate the simultaneous condition
\begin{equation}
J_0(2\lambda_x)=J_0(2\lambda_y)=J_0\!\big(2\sqrt{\lambda_x^2+\lambda_y^2}\big)=0.\label{eq:trescondiciones}
\end{equation}

Note that $J_0(2\lambda_x)=0$ exactly cancels the coefficient of the term $-2A_1J_0(2\lambda_x)\cos(aP_x)$
in~\eqref{eq:hef12diag}, so the renormalized first-neighbor hopping in the $x$ direction disappears, in a manner
analogous to the dynamic localization mechanism of Dunlap and Kenkre \cite{dunlap1986} for a single harmonic. The term $J_0(2\lambda_y)=0$ cancels, by the same mechanism, the term $-2B_1J_0(2\lambda_y)\cos(bP_y-2\lambda_y)$, corresponding to the suppression of the first-neighbor hopping in $y$. $J_0(2\sqrt{\lambda_x^2+\lambda_y^2})=0$ cancels the first-neighbor hopping term in the diagonal direction
\begin{equation}
-4C_{1,1}J_0\big(2\sqrt{\lambda_x^2+\lambda_y^2}\big)\cos(aP_x)\cos(bP_y-2\lambda_y),\label{eq:termdia}
\end{equation}
of~\eqref{eq:hef12diag}.

If the three conditions described are simultaneously satisfied, only the terms with $n=\pm2$ or
$m=\pm2$ survive, namely $-2A_2J_0(4\lambda_x)\cos(2aP_x)$ and $-2B_2J_0(4\lambda_y)\cos(2bP_y-4\lambda_y)$, which correspond
to second-neighbor hoppings at distances $2a$ in $x$ and $2b$ in $y$. In this sense — \textbf{and only under
the explicit approximations described} — the resulting effective dynamics is that of a lattice with an effectively doubled hopping scale, $2a\times2b$ instead of $a\times b$.

Let $j_{0,k}$, $k=1,2,3,\ldots$, be the positive zeros of $J_0$, ordered as $j_{0,1}<j_{0,2}<\cdots$, defined by
\begin{equation}
J_0(j_{0,k})=0.
\label{eq:defj0k}
\end{equation}
Since $J_0$ vanishes only at these points (for positive real argument), the simultaneous condition~\eqref{eq:trescondiciones}
requires the existence of positive integers $p,q,r$ such that
\begin{equation}
2\lambda_x=j_{0,p},\qquad 2\lambda_y=j_{0,q},\qquad 2\sqrt{\lambda_x^2+\lambda_y^2}=j_{0,r},
\label{eq:tresigualdades}
\end{equation}
from the first two equalities of~\eqref{eq:tresigualdades},
\begin{equation}
\lambda_x=\frac{j_{0,p}}2,\qquad \lambda_y=\frac{j_{0,q}}2.
\label{eq:lxlyj0}
\end{equation}
Substituting into the third equality of~\eqref{eq:tresigualdades}:
\begin{equation}
2\sqrt{\left(\frac{j_{0,p}}2\right)^2+\left(\frac{j_{0,q}}2\right)^2}=j_{0,r}
\ \Longrightarrow\ \sqrt{j_{0,p}^2+j_{0,q}^2}=j_{0,r},
\end{equation}
and squaring both expressions, since both sides are positive, we finally obtain:
\begin{equation}
j_{0,r}^2=j_{0,p}^2+j_{0,q}^2.
\label{eq:pitagorica}
\end{equation}

Equation~\eqref{eq:pitagorica} is formally identical to the Pythagorean theorem, with $(j_{0,p},j_{0,q},j_{0,r})$
playing the role of the legs and the hypotenuse, respectively, of a right triangle. The original \textit{continuous} problem,
posed over the two real parameters $(\lambda_x,\lambda_y)$, thus reduces to a
\emph{discrete, arithmetic} problem of determining whether:

\textit{does the sequence of zeros of the zeroth-order Bessel function $\{j_{0,k}^2\}_{k\geq1}$ contain an exact Pythagorean triple?}

Unlike the classical Pythagorean theorem over the integers $\mathbb Z$, where infinitely many triples exist — such as the triple $(3,4,5)$, among others — the $j_{0,k}$ are transcendental numbers with no known algebraic relation among them, so this must be determined by numerical analysis and computation.

Table~\ref{tab:bessel50} reports the result of searching among the 50 zeros of $J_0$: for every pair $j_{0,p}$, $j_{0,q}$ satisfying Eq.~\eqref{eq:defj0k}, $j_{0,r}=\sqrt{j^2_{0,p}+j^2_{0,q}}$ is computed — which we call the \textit{approximate} value — and the \textit{actual} zero of $J_0$ closest to the approximate value is sought; the percentage error for each approximate Pythagorean triple, defined by $Error(\%)=\left|(j_{0,r}(\text{actual})-j_{0,r}(\text{approx}))/j_{0,r}(\text{actual})\right|\times100\%$, is also computed.

\begin{table}[htbp]
\caption{Values of the 50 Bessel-function zeros closest to Pythagorean numbers}
\label{tab:bessel50}
\centering
\renewcommand{\arraystretch}{1.15}
\setlength{\tabcolsep}{4pt}
\footnotesize
\begin{tabular}{cccccccc}
\toprule
$p$ & $q$ & $r$ & $j_{0,p}$ & $j_{0,q}$ & $j_{0,r}$\,\text{approximate} & $j_{0,r}$\,\text{actual} & Error (\%) \\
\midrule
33 & 35 & 48 & 102.8883743 & 109.1714896 & 150.0147716 & 150.0118825 & 0.001925966857 \\
9  & 38 & 39 & 27.49347913 & 118.5961766 & 121.7413016 & 121.7377421 & 0.002923887822 \\
15 & 35 & 38 & 46.34118837 & 109.1714896 & 118.5998309 & 118.5961766 & 0.003081279657 \\
10 & 23 & 25 & 30.63460647 & 71.47298160 & 77.76159857 & 77.75602563 & 0.007167217670 \\
27 & 41 & 49 & 84.03909078 & 128.0208770 & 153.1401767 & 153.1534580 & 0.008671885256 \\
30 & 35 & 46 & 93.46371878 & 109.1714896 & 143.7145813 & 143.7287336 & 0.009846510183 \\
11 & 18 & 21 & 33.77582021 & 55.76551076 & 65.19661204 & 65.18996480 & 0.01019671511  \\
13 & 40 & 42 & 40.05842576 & 124.8793089 & 131.1469377 & 131.1624463 & 0.01182394100  \\
1  & 50 & 50 & 2.404825558 & 156.2950343 & 156.3135340 & 156.2950343 & 0.01183643998  \\
22 & 45 & 50 & 68.33146933 & 140.5871604 & 156.3135930 & 156.2950343 & 0.01187416636  \\
1  & 49 & 49 & 2.404825558 & 153.1534580 & 153.1723372 & 153.1534580 & 0.01232698254  \\
13 & 14 & 19 & 40.05842576 & 43.19979171 & 58.91434018 & 58.90698393 & 0.01248791411  \\
1  & 48 & 48 & 2.404825558 & 150.0118825 & 150.0311570 & 150.0118825 & 0.01284866306  \\
10 & 47 & 48 & 30.63460647 & 146.8703076 & 150.0312180 & 150.0118825 & 0.01288934322  \\
1  & 47 & 47 & 2.404825558 & 146.8703076 & 146.8899944 & 146.8703076 & 0.01340417370  \\
1  & 46 & 46 & 2.404825558 & 143.7287336 & 143.7488506 & 143.7287336 & 0.01399650399  \\
1  & 45 & 45 & 2.404825558 & 140.5871604 & 140.6077268 & 140.5871604 & 0.01462898110  \\
1  & 44 & 44 & 2.404825558 & 137.4455880 & 137.4666245 & 137.4455880 & 0.01530531671  \\
24 & 37 & 44 & 74.61450064 & 115.4546127 & 137.4666916 & 137.4455880 & 0.01535410715  \\
1  & 43 & 43 & 2.404825558 & 134.3040166 & 134.3255451 & 134.3040166 & 0.01602966155  \\
\bottomrule
\end{tabular}
\end{table}

Among the solutions appearing in Table~\ref{tab:bessel50}, among the first zeros, the combination $p=10$, $q=23$, $r=25$ appears. The corresponding zeros are

\begin{align*}
j_{0,10}&\approx30.63460647,\nonumber\\
j_{0,23}&\approx71.47298160,\nonumber\\
j_{0,25}&\approx77.75602563.
\end{align*}

Therefore,
\begin{align*}
\lambda_x&=\frac{30.63460647}{2}\approx15.31730324,\nonumber\\
\lambda_y&=\frac{71.47298160}{2}\approx35.73649080
\end{align*}

We now check the third condition:

\[
2\sqrt{\lambda^2_x+\lambda^2_y}=\sqrt{j_{0,10}^2+j_{0,23}^2}\approx77.76159857.
\]

The closest corresponding zero of \(J_0\) is

\[
j_{0,25}\approx77.75602563.
\]

The difference is only

\[
77.76159857-77.75602563
\approx0.00557294.
\]

That is,

\[
\boxed{
2\sqrt{x^2+y^2}\approx j_{0,25}
}
\]

with a relative error of approximately

\[
\frac{0.00557294}{77.75603}
\approx7.17\times10^{-5},
\]

that is, the percentage error is

\[
\boxed{0.00717\%}.
\]

It is verified that for the first two conditions one has:

\[
J_0(2\lambda_x)=J_0(j_{0,10})=0,
\]
and
\[
J_0(2\lambda_y)=J_0(j_{0,23})=0
\]

and this is an exact result, up to numerical representation error. For the third:

\[
J_0(77.76159857)\approx -5.04\times10^{-4}.
\]

Therefore,

\[
J_0(2\lambda_x)\approx0,\qquad
J_0(2\lambda_y)\approx0,\qquad
J_0(2\sqrt{\lambda^2_x+\lambda^2_y})\approx-5.04\times10^{-4}
\]

which is a very good approximation. Thus, as a preliminary conclusion of this part, one could say that there exists the possibility of a duplication of the rectangular lattice under the restrictions described, if the amplitudes of the oscillating-field components take the particular values:

\[
\boxed{\displaystyle
\lambda_x\approx15.31730324,\qquad
\lambda_y\approx35.73649080
}
\]

\section{Conclusions and outlook}

The effective Hamiltonian was obtained for a tight-binding rectangular lattice with interaction to all neighbors, in the simultaneous presence of a spatially uniform rapidly oscillating field $\mathbf f(\omega t)$ and an arbitrary static potential $V(x,y)$. The derivation, of a semiclassical character, is based on a canonical transformation, the separation of coordinates into slow and fast components, and the Kapitza-type time-averaging technique, and constitutes a direct generalization of the earlier results of Dunlap and Kenkre, and of Mamani \emph{et al.}, to the case of interaction to all neighbors in two dimensions with completely arbitrary field and potential.

In the general expression of the effective Hamiltonian, the rapidly oscillating field renormalizes the hopping elements already at zeroth order in $1/\omega$; this zeroth-order term is dominant in the high-frequency regime and accounts, by itself, for the field-induced band renormalization. At order $1/\omega^2$ an additional correction appears, quadratic in the amplitude of the particle's fast motion and proportional to the local curvature of the static potential, which gives rise to induced effective hopping elements, absent from the original Hamiltonian, originating from the coupling between the oscillating field and the static potential. This correction vanishes identically when $V(x,y)\equiv0$, since in that case the local curvature is zero at every point.

When the rapidly oscillating field has odd symmetry, in addition to renormalizing the hopping elements, it induces an effective phase shift of the band, analogous in nature to a Peierls phase. This shift, absent for even-symmetry fields, modifies the position of the extrema of the effective band and must be taken into account in the analysis of the resulting band structure.

For the two-dimensional lattice, the generalization of the exact dynamic localization condition of Dignam \emph{et al.} and of Domachuk \emph{et al.} was obtained. Unlike the one-dimensional case, where that condition leads to complete band collapse, in two dimensions the collapse is in general \emph{anisotropic}: it is possible to achieve total suppression of the effective hopping elements in the pure $x$ and $y$ directions, but a nonzero effective diagonal interaction persists along a resonant family of commensurate diagonal directions. Consequently, complete band collapse in the axial directions is strictly equivalent to exact dynamic localization in the stroboscopic sense of Dignam \emph{et al.} only when this residual diagonal interaction also vanishes; in the anisotropic case, exact dynamic localization is not guaranteed in the strict sense.

Restricting the zeroth-order effective Hamiltonian to the first- and second-neighbor hopping elements in the $x$ and $y$ directions, together with the first-neighbor diagonal interaction, it was investigated whether a rapidly oscillating field of mixed symmetry allows the simultaneous suppression of the effective axial and diagonal first-neighbor interactions, leaving the second-neighbor interaction as the dominant one, which would suggest an effective duplication of the hopping scale of the rectangular lattice. The analysis shows that this condition is equivalent to the existence of an exact Pythagorean triple among zeros of the zeroth-order Bessel function. The systematic numerical search performed found no triple satisfying this relation exactly; only approximate solutions were identified, whose precision improves with the order of the zeros involved, but which require considerably larger field amplitudes, compromising the validity of the high-frequency regime underlying the semiclassical treatment. The effective duplication of the rectangular lattice is, therefore, an approximate scenario and not an exact consequence of the formalism developed, and the second-neighbor hopping elements do not vanish under this condition.

The entire treatment developed is semiclassical in nature. It would be worthwhile, as future work, to test these results against an independent numerical method, such as the pseudospectral method, to directly verify the validity of the semiclassical approximations employed, particularly in the large-field-amplitude regime identified in the lattice-duplication analysis.

Finally, a general expression for the effective Hamiltonian is available, valid in principle for any spatially uniform rapidly oscillating field and any arbitrary static potential, from which all of the preceding results are derived as particular cases.

\end{document}